\documentclass[review]{elsarticle}
\usepackage{float}
\usepackage[pagewise]{lineno}
\usepackage[dvipsnames]{xcolor}
\usepackage{framed}
\usepackage{multicol}
\usepackage{amssymb}
\usepackage{todonotes}     
\usepackage[utf8]{inputenc}
\usepackage[ruled, vlined]{algorithm2e}
\usepackage{graphicx}
\usepackage{subcaption}
\usepackage{tabularx}
\usepackage{amsmath}
\usepackage{multirow}
\usepackage{setspace}
\usepackage{tabu}
\usepackage{multicol}

\DeclareMathAlphabet{\mathcal}{OMS}{cmsy}{m}{n}
\DeclareMathAlphabet\mathbfcal{OMS}{cmsy}{b}{n}
\newcommand{\mat}[1]{\boldsymbol{#1}}

\usepackage{siunitx}
\usepackage[colorlinks, allcolors=blue]{hyperref}
\usepackage[noabbrev]{cleveref}

\usepackage[percent]{overpic}

\usepackage[normalem]{ulem}
\usepackage{color}

\journal{International Journal of Heat and Fluid Flow}
\begin{document}
\begin{frontmatter}
\title{Predicting the temporal dynamics of turbulent channels through deep learning}

\author[1,2]{Giuseppe Borrelli}
\ead{borrelli@kth.se}

\author[2]{Luca Guastoni}

\author[2]{Hamidreza Eivazi}

\author[2]{Philipp Schlatter}

\author[2]{Ricardo Vinuesa}

\address[1]{Faculty of Aerospace Engineering, Alma Mater Studiorum - University of Bologna, Forlì, Italy}
\address[2]{FLOW, Engineering Mechanics, KTH Royal Institute of Technology, Stockholm, Sweden}


\cortext[cor]{Corresponding author}

\begin{abstract}
The success of recurrent neural networks (RNNs) has been demonstrated in many applications related to turbulence, including flow control, optimization, turbulent features reproduction as well as turbulence prediction and modeling. With this study we aim to assess the capability of these networks to reproduce the temporal evolution of a minimal turbulent channel flow. We first obtain a data-driven model based on a modal decomposition in the Fourier domain (which we denote as FFT-POD) of the time series sampled from the flow. This particular case of turbulent flow allows us to accurately simulate the most relevant coherent structures close to the wall. Long-short-term-memory (LSTM) networks and a Koopman-based framework (KNF) are trained to predict the temporal dynamics of the minimal-channel-flow modes. Tests with different configurations highlight the limits of the KNF method compared to the LSTM, given the complexity of the flow under study. Long-term prediction for LSTM show excellent agreement from the statistical point of view, with errors below $2 \%$ for the best models with respect to the reference. Furthermore, the analysis of the chaotic behaviour through the use of the Lyapunov exponents and of the dynamic behaviour through Poincaré maps emphasizes the ability of the LSTM to reproduce the temporal dynamics of turbulence. Alternative reduced-order models (ROMs), based on the identification of different turbulent structures, are explored and they continue to show a good potential in predicting the temporal dynamics of the minimal channel.
\end{abstract}


\end{frontmatter}

\section{Introduction}
\label{sec:introduction}
Turbulence is undeniably one of the most fascinating and complex phenomena in nature, and it has attracted significant efforts of the scientific community in the past 150 years. The researches in this area have a great relevance both from the scientific perspective as well as for their applicability in numerous technological fields. 
Nowadays, it is still not possible to define a theoretical model able to characterize all the structures and mechanisms which come into play when turbulence occurs. The complex nature of this phenomenon has been traditionally studied either through experiments or employing numerical simulations. 
Recently, machine learning (ML) has offered a third option to enrich the knowledge we have about this subject, also thanks to the development of more powerful deep neural networks (DNNs) over the last years. 
Some examples include improved modelling results for Reynold--averaged Navier--Stokes (RANS)~\citep{vinuesa_role_of_AI_SDG} and large-eddy simulations (LESs), flow predictions~\citep{kutz_deep,jimenez_machine,duraisamy_modeling,brunton_Machine,Jiang_datadriven_turb_modeling,guastoni_CNN}, flow control and optimization strategies~\citep{rabault_artificial,raibaudo_machine,vinuesa_flowcontrol_wings}, generation of inflow conditions~\citep{fukami_inflow}, extraction of flow patterns~\citep{raissi_hidden,eivazi_modal_decomp,eivazi_Physics_informed_RANS}, machine-learning-based reduced-order models~\citep{nakamura_CNN_LSTM,vinuesa_ML_turbulence} and prediction of the temporal dynamics~\citep{srinivasan,Eivazi_koopman}. 
The capability of a network to predict the temporal evolution of a turbulent flow is the focus of this study. One of the advantages of achieving this lies in the possibility of generating turbulent datasets employing a limited amount of initial conditions. Successful results have been obtained for a low-order nine-equation model~\citep{mohelis}, both for the statistics and the dynamical behaviour, employing models such as the multi-layer perceptron (MLP), long-short-term-memory (LSTM) network and Koopman with non-linear forcing operator (KNF)~\citep{srinivasan,Eivazi_koopman}. 

When applying machine learning to fluid dynamics it is important to understand if ML methods are more convenient when compared to other classical approaches, \emph{i.e.} if they are more accurate and efficient. Moreover, the knowledge we have about the flow has to be embedded in the models in order to improve the efficiency and the accuracy of the training and the predictions~\citep{vinuesa_ML_turbulence}.
Finally, we have to account for some possible ML alternatives which might be more appropriate for the solution of some problems, as shown in studies such as the one from \cite{Eivazi_koopman}, where the KNF outperformed the deep-learning models (LSTM and MLP) in the prediction of the temporal dynamics of non-linear systems. 

Our purpose is to employ a data-driven framework in order to have a model which is originated by the turbulence data, to overcome the limitations associated with other previous studies. Given this data-driven model, our aim is to assess the capability of some of the aforementioned architectures to predict the temporal evolution of the minimal channel.
The possibility of performing predictions on a dataset derived from a minimal channel has already been investigated by \cite{nakamura_CNN_LSTM}. In their case, the model reduction is achieved through a convolutional neural-network auto-encoder (CNN--AE) and the training and prediction of the temporal coefficients in the low-dimensional space is handled by the LSTM network. Our model is defined, instead, with an energy-based truncation derived from the outputs of a proper-orthogonal decomposition (POD), first introduced by \cite{lumley_POD} and widely employed to achieve modal decomposition in the context of turbulent flows. We focus on a specific type of POD which is performed on the Fourier transform of the velocity fields~\citep{webber_sirovich}, which is more convenient from the computational standpoint, compared with the standard POD, usually performed in the physical domain. 

The article is structured as follows: in $\S$\ref{sec:computational_framework} simulation and the data-driven model are introduced; in $\S$\ref{sec:LSTM_pred} the theory related to recurrent neural networks (RNNs) is presented and the results for the LSTM network are discussed; $\S$\ref{sec:KNF_pred} gives an overview about the Koopman-based framework and highlights the limitations of this method applied to the data-driven case; in section $\S$\ref{sec:ROMs} the possibility of implementing reduced-order models is explored; and, finally, a summary and a discussion of the results are provided in  $\S$\ref{sec:discussion}.

\section{Computational framework}
\label{sec:computational_framework}

\subsection{Numerical simulations}
\begin{figure*}[h]
    \centering
	\includegraphics[width=3.5in]{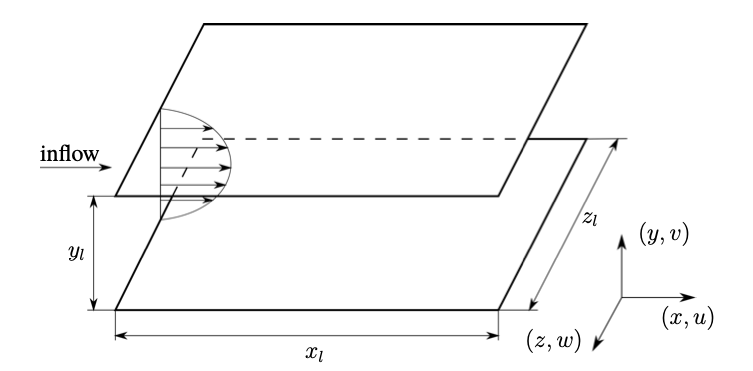}
	\caption{Schematic representation of a channel flow.}
	\label{channel_flow}
\end{figure*}

We carry out a direct numerical simulation (DNS) of the minimal channel following the parameters described in the work by \cite{jimenez_moin}. The channel flow is schematically represented in \cref{channel_flow} and the box size is set to $x_l = 0.6 \pi h$, $y_l = 2h$ and $z_l = 0.18 \pi h$, where $h = 1$ denotes the half--height of the channel.  
The resolution is $32 \times 129 \times 16$, representing the number of grid points in the streamwise (\emph{x}), wall-normal (\emph{y}) and spanwise (\emph{z}) directions, respectively. We introduce the reference velocity $U_{{\rm cl}}$ which corresponds to the centerline velocity of a laminar parabolic profile with the same volume flux as the one of the minimal channel flow. We will scale all the lengths and the velocities of the paper with respect to $h$ and $U_{{\rm cl}}$, respectively. The laminar centerline-velocity based Reynolds number is set to $Re_{{\rm cl}}=U_{{\rm cl}} h / \nu = 5,000$, which leads to a friction Reynolds number $Re_{\tau}=h u_{\tau}/\nu=202$ (where $\nu$ is the fluid kinematic viscosity and $u_{\tau}$ is the friction velocity). 
The minimal channel flow was simulated with the Fourier--Chebyshev numerical code SIMSON~\citep{simson_manual}, a spectral solver which enables an efficient solution of the Navier--Stokes equations for some canonical cases.
A simulation initialized with a superimposed random noise was performed from time $t = 0$ to $t = 10,000$, in order to have a fully-developed turbulent flow. Discarding this initial transient, the condition at $t = 10,000$ is employed as the initial one to generate the turbulent database composed by $800,000$ snapshots on a time span of $160,000$ time units employing a constant time step for the sampling of $\Delta t_{\rm s} = 0.2$.
A second-order Crank--Nicholson (CN) scheme is employed to model the linear terms of the Navier--Stokes equations, whilst the non-linear ones  employ a third-order four-stages Runge--Kutta (RK3) scheme. The flow is driven by a streamwise pressure gradient, which is adjusted at each time step to maintain a constant flow rate. 

\subsection{Data-driven model}
The data-driven model is constructed based on the outputs obtained after performing the proper-orthogonal decomposition (POD). An optimal low-order reconstruction of the flow from the energetic point of view is achievable by defining a truncation, since the contribution of the modes is rearranged with decreasing energy. The method is based on the Karhunen--Loéve theorem~\citep{berkooz_pod}, which allows identifying a deterministic basis and a stochastic signal, given a number of turbulent velocity fields (snapshots). The stochastic part is described as:
\begin{equation}
     \mathbf{u}(\mathbf{x},t) = \sum_{j=1}^{N_p} a_j(t) \mathbf{\Phi}_j(\mathbf{x}),
    \label{KLdecomp}
\end{equation}
and defines the modal decomposition which separates the spatial modes $\mathbf{\Phi}_j(\mathbf{x})$ from the associated temporal coefficients $a_j(t)$.
Before performing the modal decomposition, the snapshots describing the evolution in time of the three-dimensional velocity fields are rearranged into a snapshot matrix $\mathbf{u}_{\rm POD}$ as follows:
\begin{equation}
    \mathbf{u}_{\rm POD}=\begin{bmatrix} 
    u_{x_1}^{t_1} & \dots  & u_{x_1}^{t_{N_t}}\\
    \vdots         & \ddots & \vdots \\
    u_{x_{N_p}}^{t_1} & \dots  & u_{x_{N_p}}^{t_{N_t}}
    \end{bmatrix},
    \label{snap}
\end{equation}
where $N_p$ is the number of grid points and $N_t$ the total number of snapshots which are collected.
In our study we focus on the streamwise component of the velocity only, \emph{i.e.} the $u$-component. This is a reasonable simplification, due to the fact that this is the main velocity-fluctuation component in channels.

Having $x$ and $z$ as homogeneous directions in the minimal channel, the POD analysis can be performed directly on the Fourier transform of the field in those directions~\citep{webber_sirovich}, \emph{i.e} we move from $\mathbf{u}(x,y,z,t)$ to $\hat{\mathbf{u}}(m,y,n,t)$, where $(m,n)$ represents the wavenumber pair which, together with the $y$-dynamics ($k$), fully specifies the mode. Given a wavenumber pair, the modal decomposition is obtained by either solving the eigenvalue problem or employing the singular-value decomposition (SVD). 
In this work we employ the SVD approach, which decomposes the snapshot matrix as follows:
\begin{equation}
    \mathbf{u}_{\rm POD} = \mathbf{U}(\textbf{x}) \cdot \mathbf{s} \cdot \mathbf{V}(t)^{\mathsf{H}},
    \label{fftpod}
\end{equation}
where superscript $\mathsf{H}$ denotes the Hermitian transpose of the matrix, $\mathbf{U}(\textbf{x})$ represents the spatial basis, $\mathbf{V}(t)$ the temporal modes and $\mathbf{s}$ the singular values, \emph{i.e.} the energy associated with each mode. The sizes of the resulting matrices are:
$\mathbf{u}_{\rm{POD}}(\mathbf{x},t) \in \mathbb{C}^{(n_y \times N_t \times n_x n_z)}$, $\mathbf{U}(\mathbf{x}) \in \mathbb{C}^{(n_y \times e \times n_x n_z)}$, $\mathbf{s} \in \mathrm{I\!R}^{(e \times e \times n_x n_z)}$ and $\mathbf{V}(t) \in \mathbb{C}^{(N_t \times e \times n_x n_z)}$, with $e = \rm{min}$ $(N_t,n_y)$ being the size associated with the economy-SVD (and is equal to $n_y$ for our case). Once the outputs are derived, the visualization of a single spatial mode, $\mathbfcal{U}_{(m,n),k} \in \mathbb{C}^{(n_x \times n_y \times n_z)}$, is obtained employing a frequency filter, $\mathbfcal{N} \in \mathrm{I\!R}^{(n_x \times n_y \times n_z)}$, over the matrix $\mathbf{U}$~\citep{webber_sirovich}. The filtering operation is performed for each quantum number $k$:
\begin{equation}
    \mathbfcal{U}_{(m,n),k} = \mathbf{U}_{k} \otimes \mathbfcal{N}_{(m,n)},
    \label{podd}
\end{equation}
where the operation $\otimes$ represents the element-wise product between matrices. To visualize the mode in the physical domain an inverse fast Fourier transform is applied on $\mathbfcal{U}_{(m,n),k}$.
It is now possible to reconstruct the velocity field for a given truncation choice, \emph{i.e.} up the the $M$-th most energetic mode:
\begin{equation}
    \mathbf{u}(\mathbf{x},t) = \sum_{i = 0}^{M} \bigg( \mathrm{Re} \{a_i(t)\cdot \mathbfcal{U}_i \} - \mathrm{Im}\{ a_i(t)\cdot \mathbfcal{U}_i \} \bigg),
    \label{FFTrecon}
\end{equation}
having $a(t) = \mathbf{s}^{\mathbf{k}} \cdot \mathbf{V}^{\mathbf{k}}(t)^{\mathsf{H}}$ as the temporal coefficient calculated after the FFT-POD method. The subscript $i$ accounts for the mode represented by a triplet $\mathbf{k} = (m,n,k)$. The imaginary contribution of the mode is subtracted when reconstructing the field, otherwise a shift with respect to the original field is observed.

\subsubsection{Energy-based truncation}
\begin{figure}[h]
    \centering
    \includegraphics[width=2.3in]{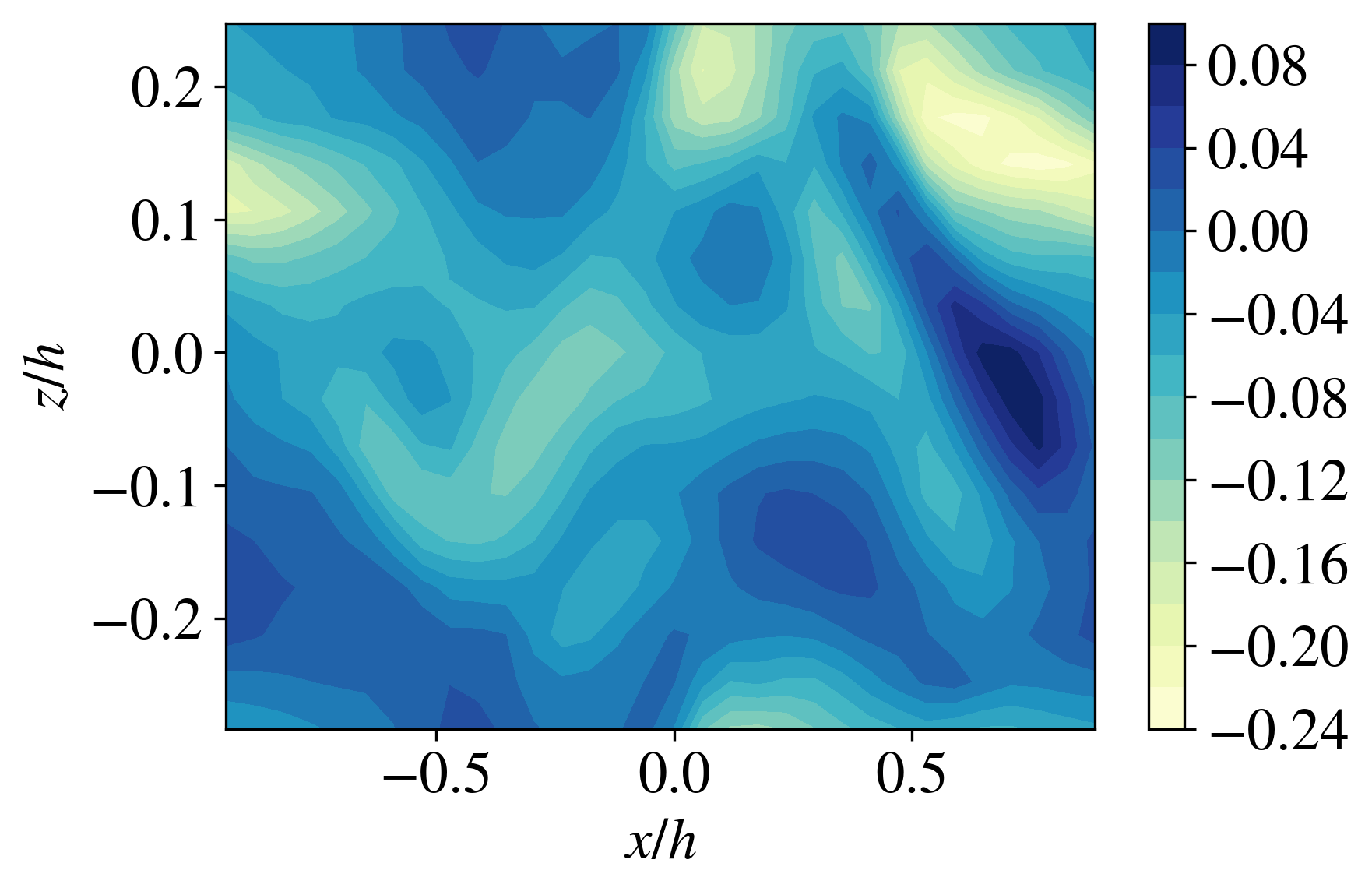}
    \includegraphics[width=2.3in]{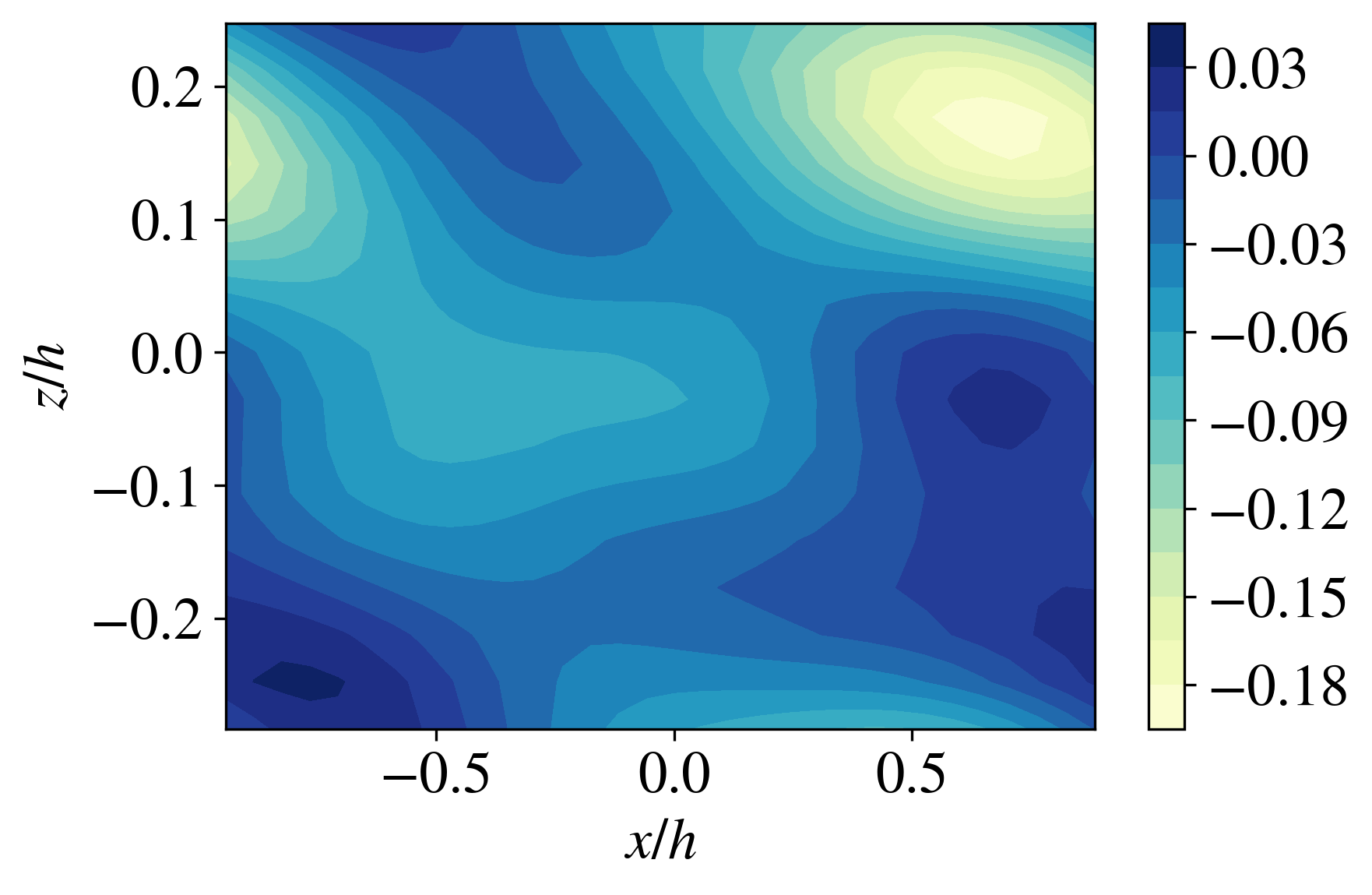}
    \includegraphics[width=2.3in]{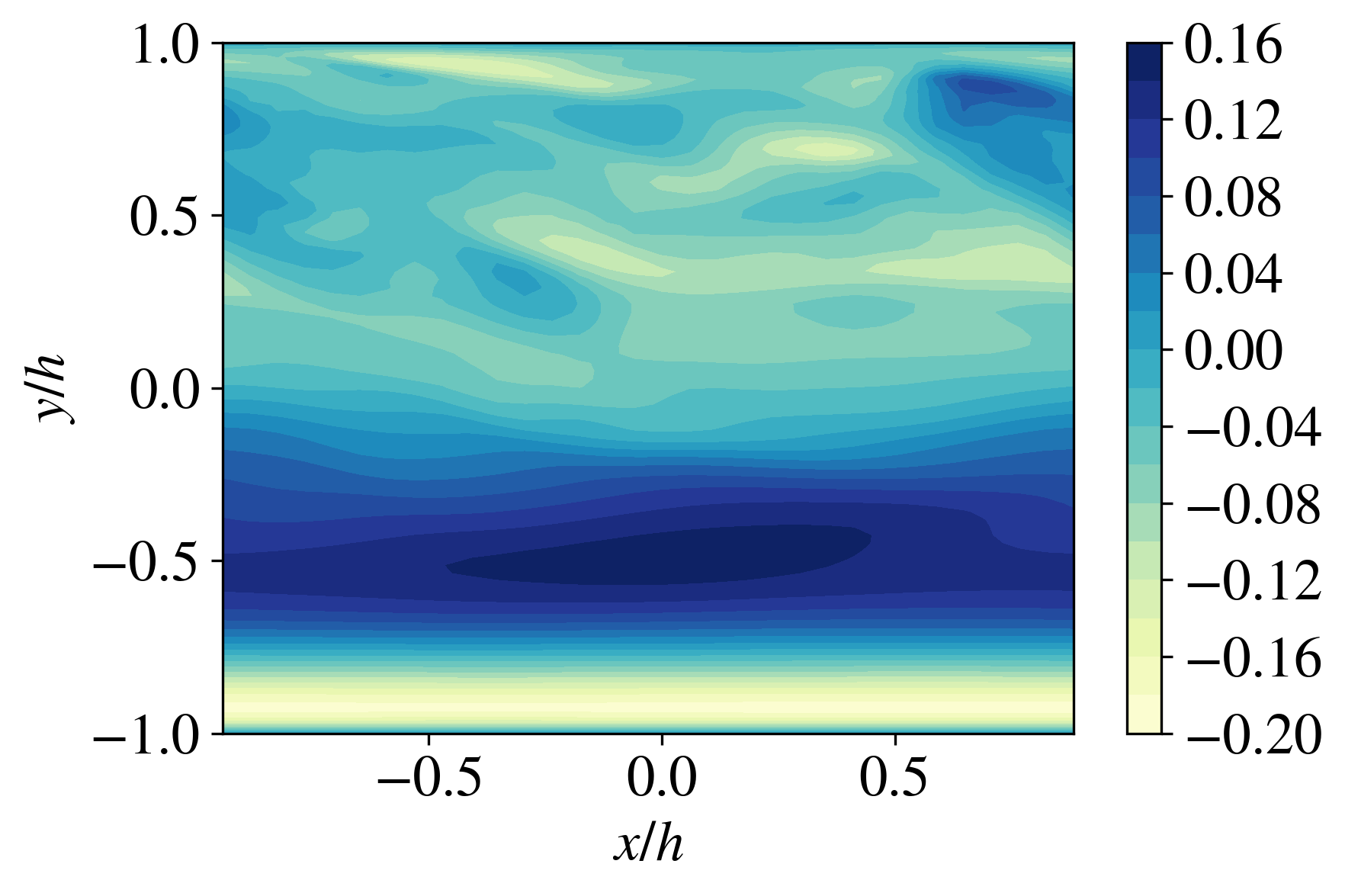}
    \includegraphics[width=2.3in]{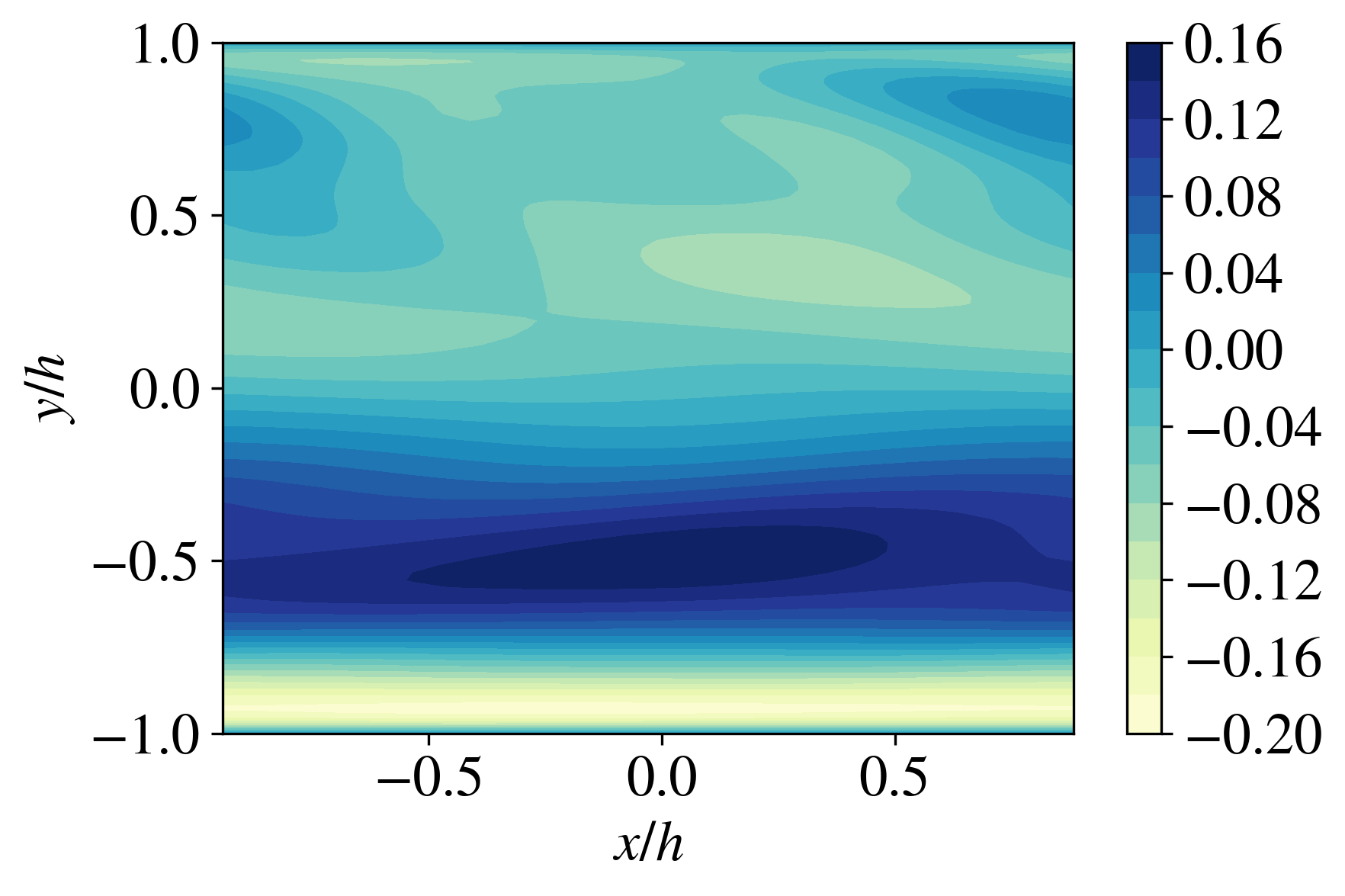}
    \caption{Visualization of the streamwise fluctuations of the velocity (\emph{left}) compared to the reconstruction which employs the 100 most energetic modes (\emph{right}). The $xz$-plane is shown above ($y^+ \approx 24$) and the $xy$-plane is shown below ($z=0$). The example reports the reconstruction at $t=10,000$ for a dataset composed by $10,000$ snapshots, where the initial transient has been discarded.}
    \label{trunc_visual}
\end{figure}

We define our model by carrying out a truncation which employs the first 100 most energetic modes. These modes capture $78.8 \%$ of the total energy related to the streamwise fluctuations and generate an error on the reconstruction of $22.2 \%$, where this error is defined as:
\begin{equation}
    \frac{||u_{\rm{orig}}-u_{\rm{recon}}||_2}{||u_{\rm{orig}}||_2},
\end{equation}
where $u_{\rm{orig}}$ are the streamwise fluctuations of the velocity reconstructed with all the modes; and $u_{\rm{recon}}$ are the ones reconstructed with a reduced amount of modes.
An example of how the velocity fields are reconstructed is shown in \cref{trunc_visual}.
The plots show the two-dimensional comparison for the $xy$- and $xz$-planes at $z=0$ and $y^+ \approx 24$, respectively, where the distance from the wall is given in wall units $y^+ = y \cdot u_{\tau}/\nu$.
Even if a difference due to the truncation is observed, the obtained model is representative of the turbulence in the minimal channel, \emph{i.e.} it reconstructs a significant percentage of the total energy and it includes the most relevant turbulent features present in channel flows.
These structures are recognizable in the modes which define the spatial basis of our model.
\begin{figure} [h]
	\centering
    \begin{overpic}
    [scale=.22,percent]{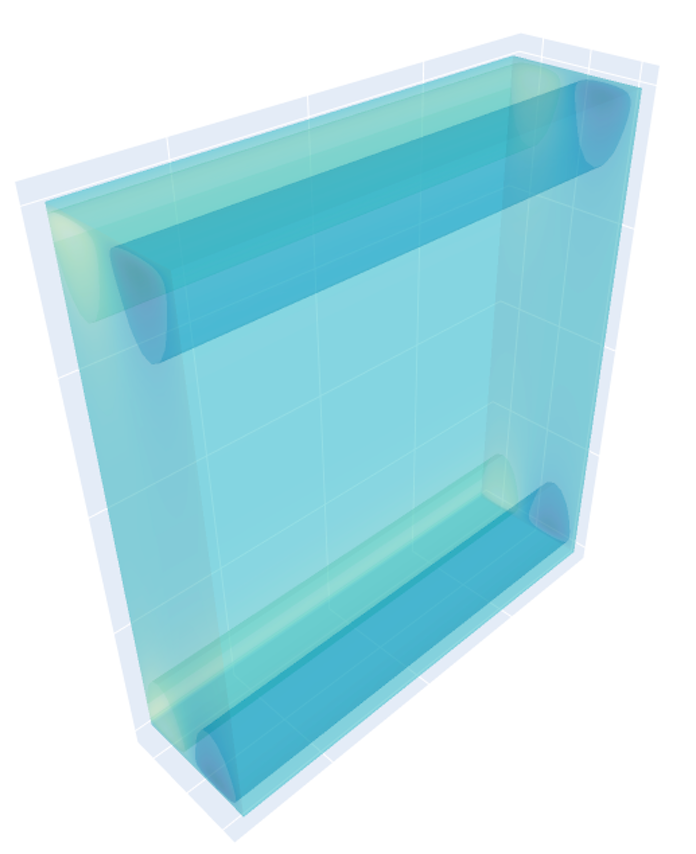} 
    \put(3,40){\textcolor{gray}{$y$}}
    \put(15,5){\textcolor{gray}{$z$}}
    \put(51,17){\textcolor{gray}{$x$}}
    \end{overpic}
    \begin{overpic}
    [scale=.22,percent]{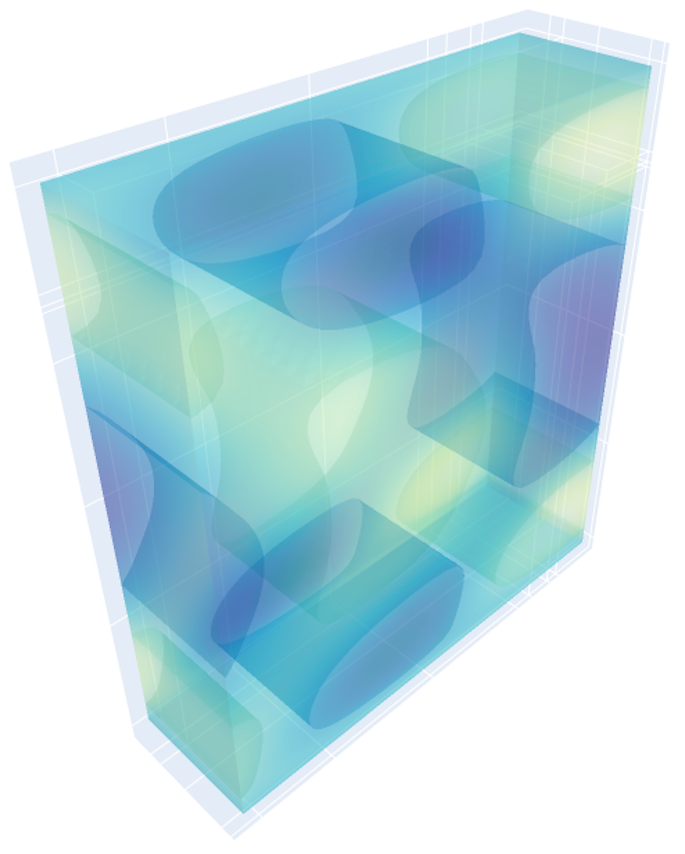} 
    \put(3,40){\textcolor{gray}{$y$}}
    \put(15,5){\textcolor{gray}{$z$}}
    \put(51,17){\textcolor{gray}{$x$}}
    \end{overpic}
    \begin{overpic}
    [scale=.22,percent]{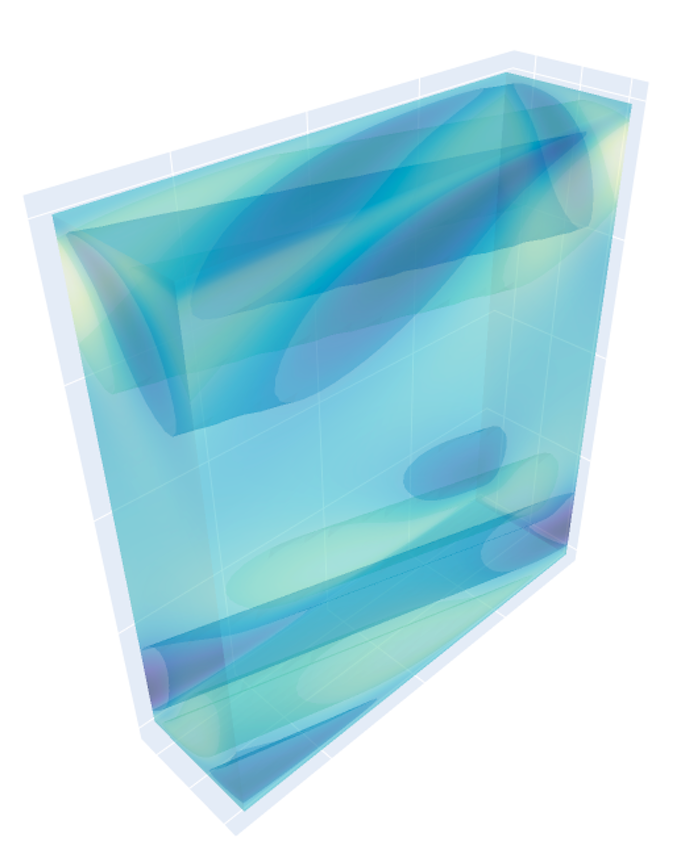} 
    \put(4,40){\textcolor{gray}{$y$}}
    \put(15,5){\textcolor{gray}{$z$}}
    \put(51,17){\textcolor{gray}{$x$}}
    \end{overpic}
    \caption{Visualization of roll mode (\emph{left}), streamwise mode (\emph{center}) and propagating mode (\emph{right}).}
    \label{turb_features}
\end{figure}
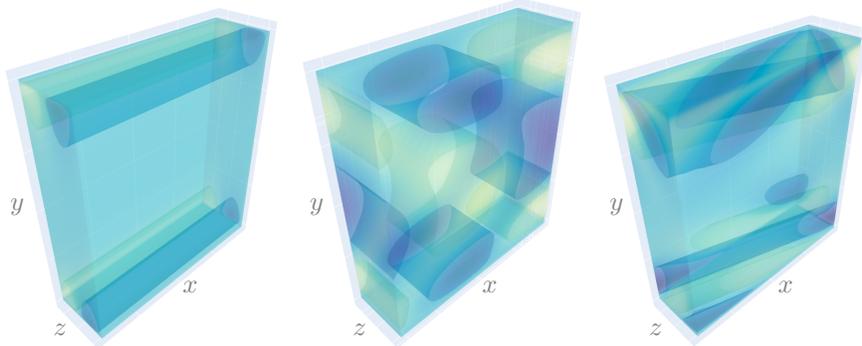
\Cref{turb_features} shows the roll modes ($m=0$), which are usually associated with the higher energy content and represent counter-rotating vortex which move the flow from the wall to the outer region and vice-versa; the streamwise modes ($n=0$) show a dependence only in the $x$-direction; and the propagating modes, described as tilted vortex with respect to the wall and with development in the three directions~\citep{webber_sirovich}. \Cref{MCF_enfrac_table} reports the energy percentage associated with the first modes.
\begin{table}[h]
\centering
	\begin{tabular}{ |c|c c c c c c c| }
	\hline
	$(m,n,k)$ & (0,1,1) & (0,1,2) & (0,0,1) & (0,0,2) & (1,1,1) & (0,1,3) & (0,1,4) \\
    \hline
    energy $\%$ & 12.69 & 11.67 & 9.30 & 5.89 & 3.82 & 3.18 & 3.00 \\
	\hline
	\end{tabular}
	\caption{Energy percentage associated with the first 7 modes present in the minimal channel.}
	\label{MCF_enfrac_table}
\end{table}

The procedure through which the fields are simulated accounts separately for the wavenumbers with opposite sign in the homogeneous directions. These exhibit the same energy content and for this reason the concept of degeneracy is applied, \emph{i.e.} these modes are considered as a single one and their energy is summed up together.

One final aspect related to the minimal channel arises from \cref{trunc_visual}, \emph{i.e.} the fact that, alternatively, the flow can be turbulent at one wall and (nearly) laminar at the other, leading to a different mean velocity profile in time. This issue was addressed by \cite{jimenez_moin}, who recognized a slow relaxation time for the statistics. For this reason, the comparison between true and predicted statistics will be justified only when employing the same time horizon and when starting from the same time instant.

\subsubsection{Dataset analysis}

After fixing the spatial basis, we focus on the analysis of the temporal modes. These modes define the dataset employed for the training and the ones that are predicted. We introduce an additional simplification by separating the fluctuations of the temporal coefficients related to the first wavenumber pair $a_{(0,0)}(t)$ and the ones associated with the other wavenumbers $a'(t)$. Doing so, we exclude the net-flux modes, which vary only along the wall-normal direction, and we are able to represent $97 \%$ of the total energy associated with $a'(t)$.
Moreover, a scaling of the dataset is necessary in order to have a correct training of the network, otherwise the gradients which enter the computations of the cost function will be too small and it will be impossible to generate significant updates of the parameters of the network. The evolution of each scaled temporal mode $\widehat{a}_j(t)$ is given by:
\begin{equation}
    \widehat{a}_j(t) = \frac{a_j(t)-\langle{a_j(t)}\rangle}{\sigma [a_j(t)]},
    \label{data_scaling}
\end{equation}
where the scaling is performed by subtracting the mean, denoted by $\langle \cdot \rangle$, and dividing everything by the variance, $\sigma$, of the signal.

\begin{figure}[h]
    \centering
    \includegraphics[width=2.3in]{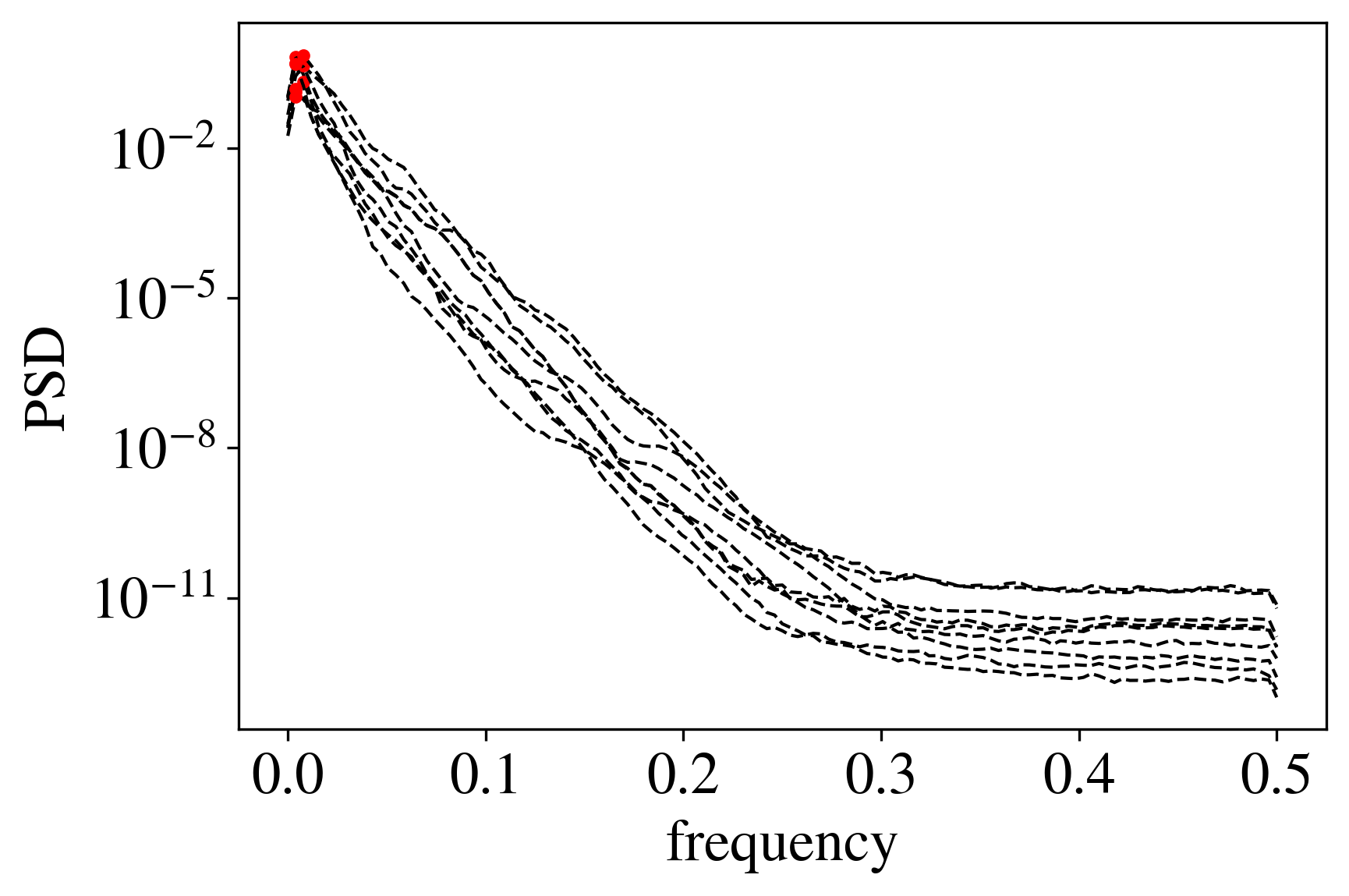}
    \includegraphics[width=2.3in]{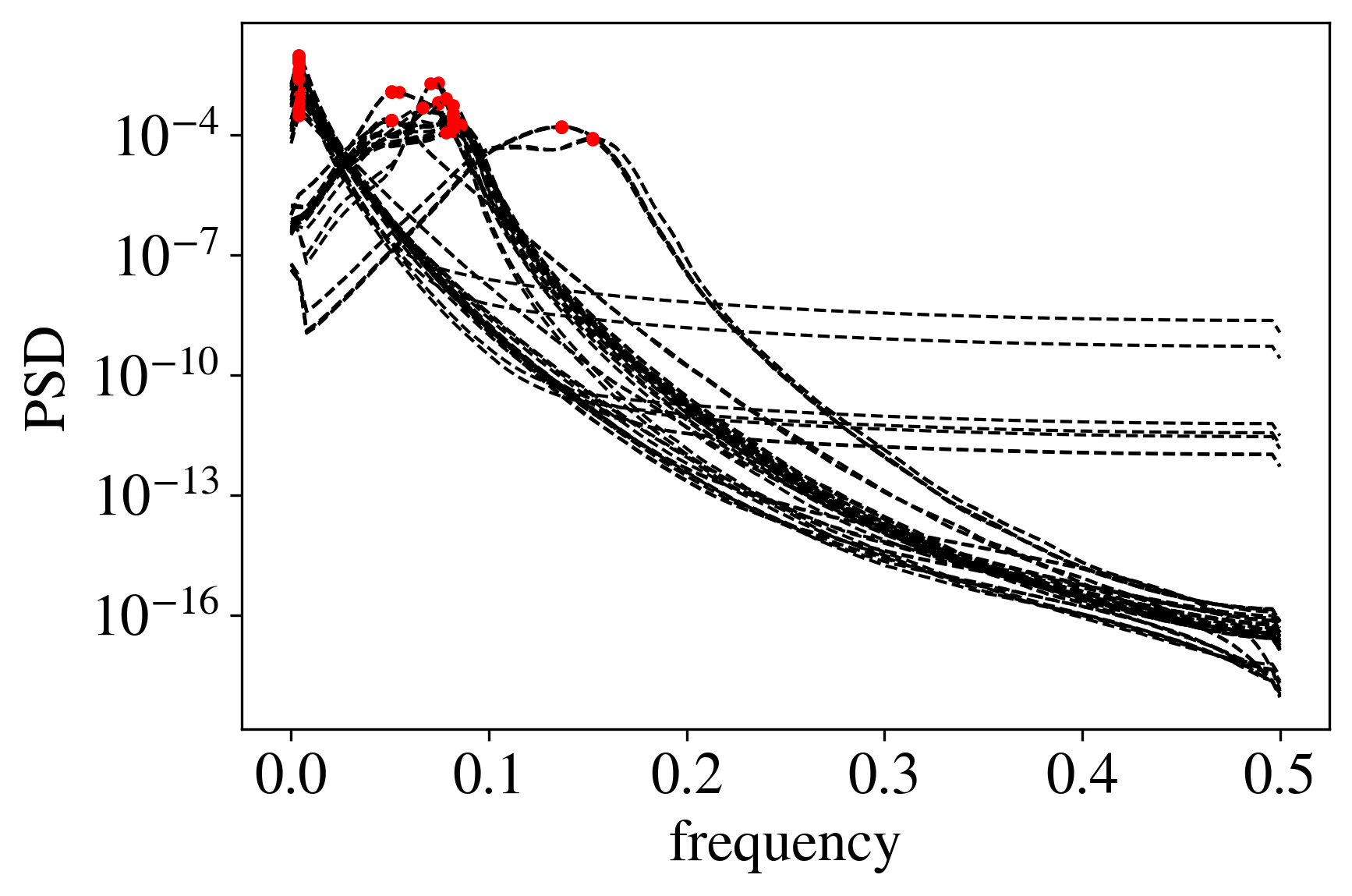}
    \caption{Comparison of the frequency content between the model by~\cite{mohelis} (\emph{left}) and the minimal channel (\emph{right}) studied here. Peaks are highlighted.}
    \label{freq_MCFvsMohelis}
\end{figure}

A difference in the frequency content of the temporal modes of the minimal channel is observed, as highlighted in \cref{freq_MCFvsMohelis}.
This feature was not reported in the studies predicting the nine-equation model~\citep{srinivasan,Eivazi_koopman}, where all the modes have similar frequency spectra.
A multi-step concept is applied to correctly capture the dynamics for each mode and is achieved by adjusting the sampling rate, employing a larger $\Delta t$ for the signals with a lower frequency and a smaller one as the high-frequency content becomes more prominent. The main idea of this approach is to have different networks which are responsible for the prediction of a reduced set of modes.

The possibility of applying the multi-step model relies on a correlation analysis performed on the signals of the groups which are identified. If the correlation between signals of different groups is low compared with that of signals from the same group then it is reasonable to use this subdivision. In our case we have low- and high-frequency signals, denoted by $[\mathbf{1}]$ and $[\mathbf{2}]$ respectively. The correlation factor, defined as $\sigma [\mathbfcal{C}_{(i,j)}] - \langle \mathbfcal{C}_{(i,j)} \rangle$ (where $\mathbfcal{C}_{(i,j)}$ is the correlation vector between the signals related to modes $i$ and $j$), belongs to the interval $[3 \times 10^{-4}, 2 \times 10^{-3}]$ for signals of the different groups, while it belongs to $[4 \times 10^{-3}, 3 \times 10^{-1}]$ for signals of the same group, leading to the conclusion that it is safe to train the networks separately. We employ a sampling rate $\Delta t_{[\mathbf{1}]} = 0.8$ for the low-frequency signals and $\Delta t_{[\mathbf{2}]} = 0.4$ for the high-frequency ones.

\section{Predictions with long-short-term-memory (LSTM) network}
\label{sec:LSTM_pred}
The success of recurrent neural networks (RNNs) to achieve the training on sequential data has been demonstrated with respect to the much simpler architecture of the multi-layer perceptron (MLP) which, instead, is limited to a point-to-point prediction, without taking into account the temporal dependencies of the input. The $input \rightarrow hidden \rightarrow output$ layers structure is preserved but, with respect to MLPs, RNNs introduce a feedback loop in the hidden layer (or layers), allowing to combine the information about the internal state of each neuron (also known as memory) with the input vector, to generate the output at a given time. Thanks to the feedback mechanism, this output becomes the input of the same neuron at the next time step and, eventually, it allows the network to learn the dynamics of the system. 
The training of the network gets more difficult as the input sequence spans a longer range in time and leads to the problem of vanishing gradients. This arises when, for each iteration, the corrections to the weights associated to the back propagation algorithm reduce the gradient until the network is not capable of learning any further.

Introduced by \cite{hochreiter_LSTM}, long-short-term-memory (LSTM) networks are ideal for long-term dependencies, such as the ones encountered in turbulent flows, thanks to the ability to control the dynamics of the recurrent connections in time by means of gating mechanisms. \Cref{LSTM_alg} shows the procedure to compute the output of the cell for each time step.
\begin{algorithm}[h]
\DontPrintSemicolon
\KwIn{Sequence $\pmb{\chi}_{1}, \pmb{\chi}_{2}, \dots \pmb{\chi}_{p}$}
\KwOut{Sequence $\pmb{\zeta}_{1}, \pmb{\zeta}_{2}, \dots \pmb{\zeta}_{p}$}
 set $\mathbf{h}_0 \leftarrow 0$\;
 set $\mathbf{C}_0 \leftarrow 0$\;
 \For{$t\leftarrow 1$ \KwTo $p$}{
  \ \ $\mathbf{f}_t \leftarrow \sigma(\mathbf{W}_{f} [\pmb{\chi}_{t}, \pmb{\zeta}_{t-1}] + \mathbf{b}_{f})$\;
  \ \ $\mathbf{i}_t \leftarrow \sigma(\mathbf{W}_{i} [\pmb{\chi}_{t}, \pmb{\zeta}_{t-1}] + \mathbf{b}_{i})$\;
  \ \ $\mathbf{\widetilde{C}}_t \leftarrow \tanh(\mathbf{W}_{f} [\pmb{\chi}_{t}, \pmb{\zeta}_{t-1}] + \mathbf{b}_{f})$\;
  \ \ $\mathbf{C}_t \leftarrow \mathbf{f}_t\otimes\mathbf{C}_{t-1} + \mathbf{i}_t\otimes\mathbf{\widetilde{C}}_t$\;
  \ \ $\mathbf{o}_t \leftarrow \sigma(\mathbf{W}_{o} [\pmb{\chi}_{t}, \pmb{\zeta}_{t-1}] + \mathbf{b}_{o})$\;
  \ \ $\pmb{\zeta}_t \leftarrow \mathbf{o}_t\otimes\tanh(\mathbf{C}_{t-1})$\;
 }
 \caption{General scheme of the LSTM algorithm.}
 \label{LSTM_alg}
\end{algorithm}
The role of each gate is as follows: the forget gate employs the current input ($\pmb{\chi}_t$) and the output of the previous time instant ($\pmb{\zeta}_{t-1}$) to define the fraction of the cell state ($\mathbf{C}_{t-1}$), which is kept in the evaluation of the current cell ($\mathbf{C}_{t}$); the input gate determines the values of the cell state to be updated (using the same quantities as the ones of the forget gate) and it computes candidates ($\Tilde{\mathbf{C}}_t$) for the update of the cell state at the present time instant; and the output gate computes the output values ($\pmb{\zeta}_i$) applying the updated cell state $\mathbf{C}_{t}$.

The LSTM network is trained over $2,000$ epochs with focus on the real part of the temporal coefficients only, having the imaginary description with a very similar dynamics. We use the hyperbolic tangent as the activation function for the internal layer, to account for the non-linearity of the problem, while the dense layer (\emph{i.e.} the one that stores the outputs) employs a linear activation function to consider the correct range of the values, due to the fact that once the dataset is scaled we are not limited to the interval $[-1,1]$ anymore (which are the limit values of the $\tanh$ function). We use a mean-squared error as a loss function and the Adam optimizer to control the evolution of the learning rate~\citep{Adam_kingma}. Albeit the Adam algorithm itself employs a correction of the learning rate during the training, an exponential decay is introduced to have a further reduction of the losses:
\begin{equation}
    {\rm LR} = {\rm LR}_0 \cdot \alpha_{\rm D}^{n_{\rm C}/n_{\rm D}} = 0.001 \cdot 0.96^{n_{\rm C}/n_{\rm D}},
    \label{LRSchedule}
\end{equation}
where ${\rm LR_0}$ is the initial learning rate, $\alpha_{\rm D}$ is the decay rate and $n_{\rm D}$ is the decay step to be considered with respect to the current step $n_{\rm C}$.
Finally, the dimension of the batch size is constant and equal to 32. This means that every 32 samples an update of the weights is made. The dataset is divided into training and validation sets: $80 \%$ of the data is employed for training and $20 \%$ for the validation. The model which gives the best loss is stored as the training advances.

\subsection{Temporal predictions and statistics}
The idea behind the prediction is to select an initial sequence of length $p$ and predict the value at $p+1$. Moving one step forward with the predictions, the $p+2$ value is then predicted, still using the $p$ values before, where, this time, one of the value of the sequence is predicted. This explains how, as we advance in time, the error related to the predictions accumulates, leading to a point-by-point prediction which is not exact. Validation losses give an indication of how accurate the predicted values are with respect to the true ones.

\begin{figure}[h]
    \centering
    \includegraphics[width=3.5in]{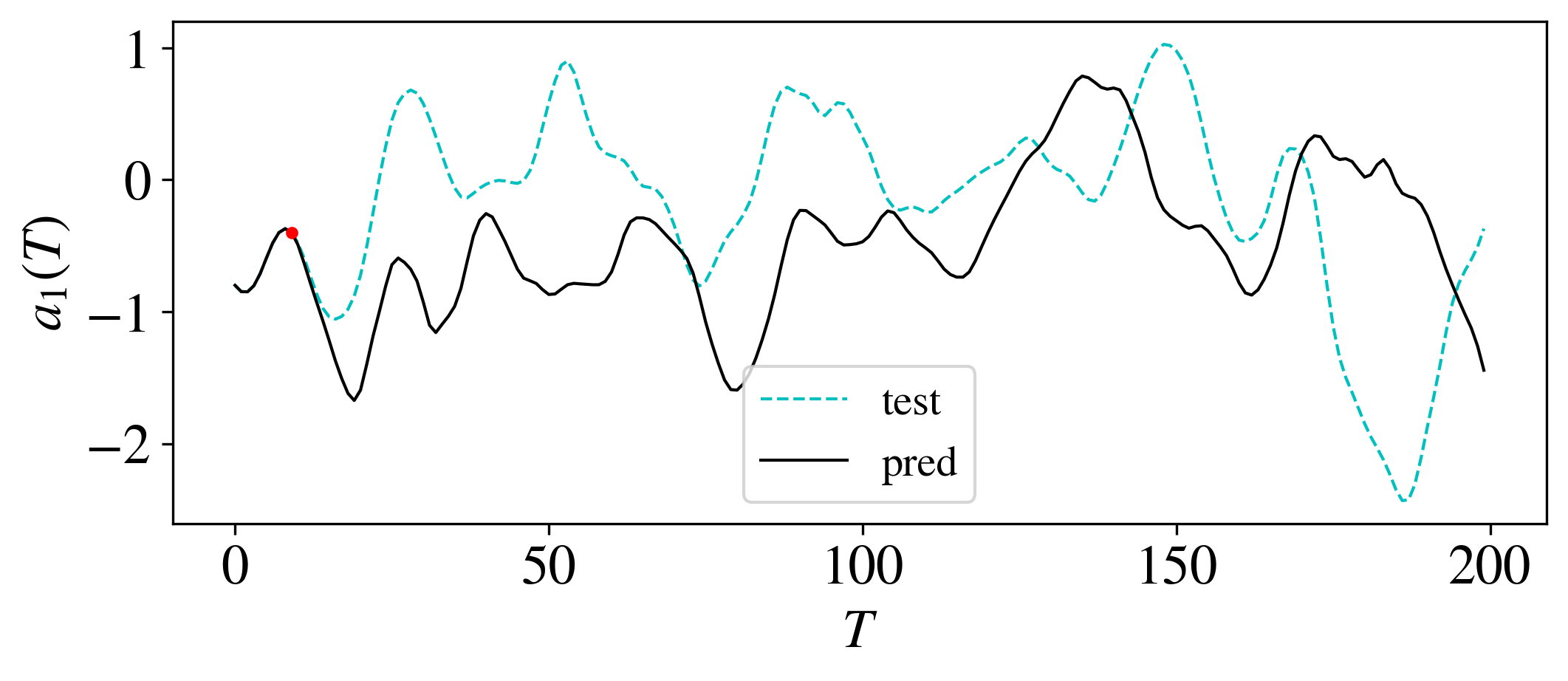}
    \includegraphics[width=3.5in]{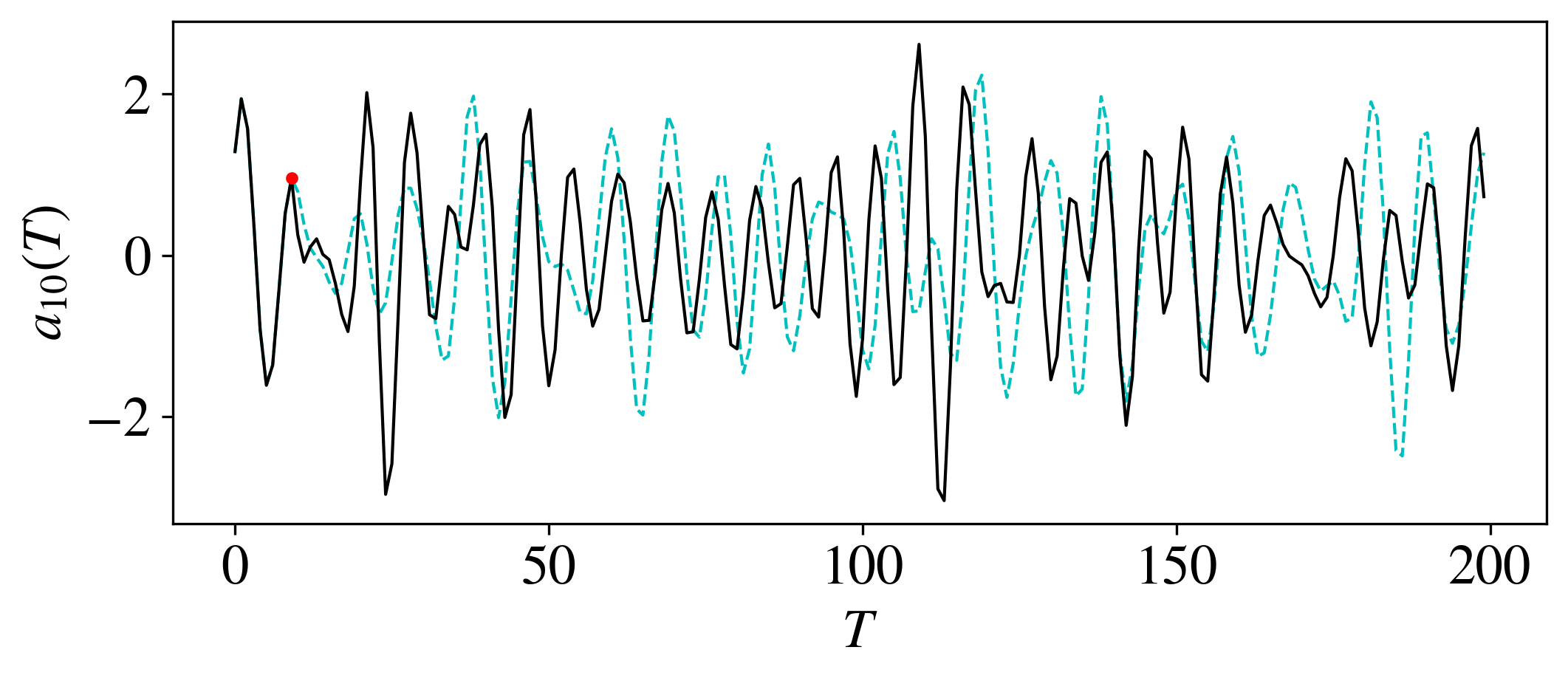}
    \caption{Short-term predictions for mode 1 (\emph{top}) and mode 10 (\emph{below}). These two modes are representative of low-frequency and high-frequency signals, respectively. The red dot represents the point where predictions start. Note that indexing starts with mode 0, so mode 1 is the second most energetic mode.}
    \label{LSTM_pred_short}
\end{figure}

\begin{figure}[h]
    \centering
    \includegraphics[width=3.5in]{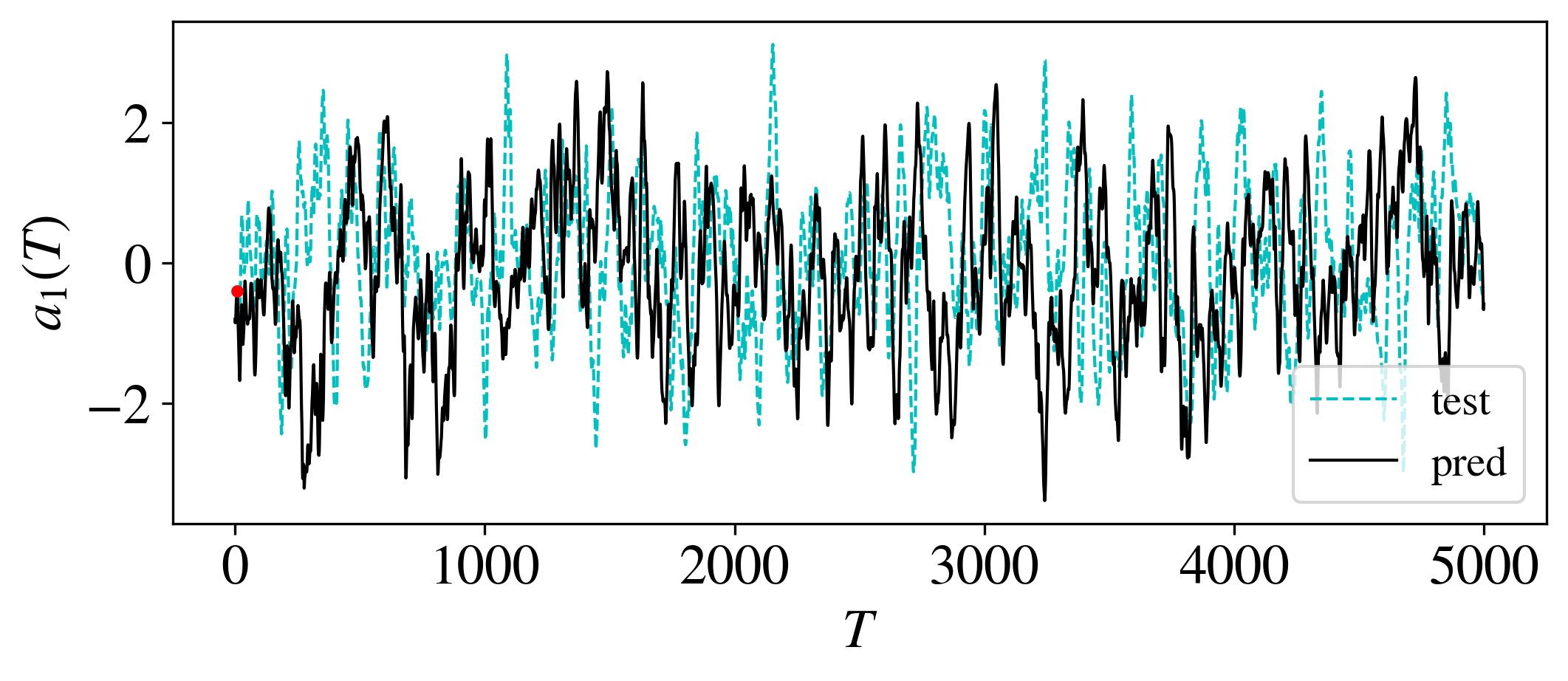}
    \includegraphics[width=3.5in]{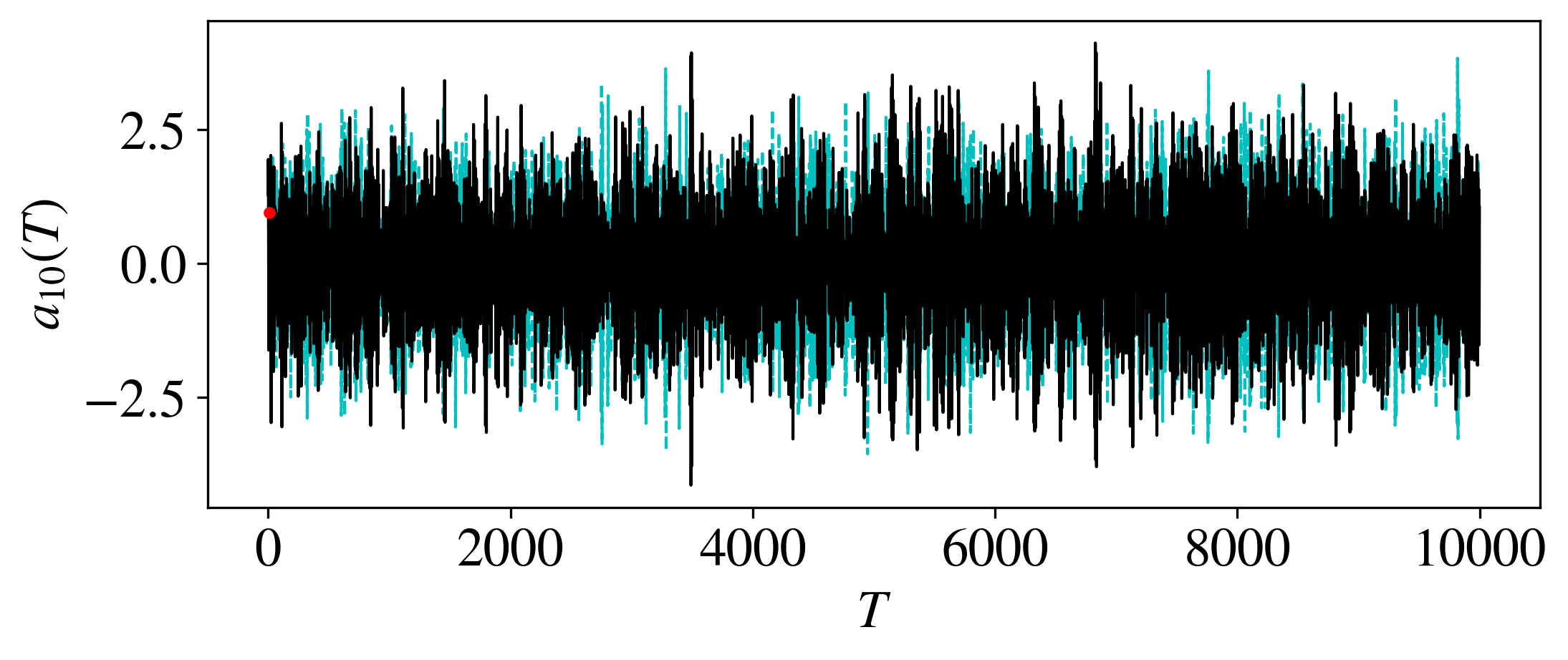}
    \caption{Long-term predictions for mode 1 (\emph{top}) and mode 10 (\emph{below}).}
    \label{LSTM_pred}
\end{figure}

The instantaneous predictions for the best model are reported in \cref{LSTM_pred_short} for the short-term horizon and in \cref{LSTM_pred} for a longer time span.
These are obtained with the LSTM-1-200-10-100 model, \emph{i.e.} a network which employs a single layer with 200 neurons, an initial vector of $p=10$ values and it has been trained with a dataset of $100,000$ snapshots.
We observe a temporal evolution which is not exactly the same as it was for the predictions based on the nine-equations model. On the other hand, the networks are still able to reproduce the frequency content of the signals and the coefficients lie in the correct range, providing a solution which is acceptable from the physical standpoint. 
The validation losses are of the order of $10^{-6}$ and $10^{-3}$ for the signals of group $[\mathbf{1}]$ and $[\mathbf{2}]$, respectively. The effect of each hyper-parameter on the statistics and on the validation is described in \cref{LSTM_stats_Hpar}.
We notice that the best models are the ones which either have a higher capacity, \emph{i.e.} more neurons are employed, or have a shorter initial vector of known coefficients. We also observe how for some cases a non-physical behaviour takes place, for instance if the number of the snapshots of the training dataset is not sufficient or when the number of cells in the hidden layer are not enough. 
This behaviour manifests itself by showing an intensification of the oscillation after a certain time horizon. 
Note that the first attempt was performed by using 90 neurons, which successfully predicted the temporal dynamics of the modes of the nine-equation model, but this model was not able to reproduce a plausible behaviour for the minimal channel.
Once the temporal evolution of the modes is predicted we can use this information together with the spatial basis in order to reconstruct the velocity fields using equation \eqref{FFTrecon} with $M=100$, after having scaled the temporal data back to the original range.
\begin{table}[h]
\centering
	\begin{tabular}{| c |  c c c |}
	\hline \textbf{LSTM-ly-cells-p-N}  & $E_{{u_{\rm{RMS}}}}[\%]$ & val. loss 1 & val. loss 2 \\
	\hline \hline
	\textit{LSTM-1-200-10-10} & 10.64 & $1.08 \times 10^{-5}$ & $1.13 \times 10^{-2}$\\
	LSTM-1-200-10-50  & 3.57 & $1.09 \times 10^{-5}$ & $7.62 \times 10^{-3}$\\
	\underline{LSTM-1-200-10-100} & \textbf{2.97} & $7.74 \times 10^{-6}$ & $\mathbf{2.03 \times 10^{-3}}$\\
	LSTM-1-200-10-200 & 6.08 & $3.97 \times 10^{-5}$ & $2.29 \times 10^{-3}$\\
	\hline
	\textit{LSTM-1-90-10-50} & 16.37 &  $1.49 \times 10^{-4}$ & $6.05 \times 10^{-3}$\\
	\underline{LSTM-1-150-10-50} & 3.52 & $2.26 \times 10^{-5}$ & $6.39 \times 10^{-3}$\\
	LSTM-1-300-10-50 & 5.67 & $\mathbf{5.01 \times 10^{-6}}$ & $7.35 \times 10^{-3}$\\
	\hline
	\underline{LSTM-1-200-5-50} & 3.28 & $9.51 \times 10^{-6}$ & $3.35 \times 10^{-3}$\\
	LSTM-1-200-20-50 & 7.85 & $2.00 \times 10^{-5}$ & $1.10 \times 10^{-2}$\\
	LSTM-1-200-40-50 & 6.86 & $1.63 \times 10^{-5}$ & $1.37 \times 10^{-2}$\\
	\hline
	\textit{LSTM-2-200-10-50} & 14.32 &  $6.47 \times 10^{-6}$ & $9.30 \times 10^{-3}$\\
	\hline
	\end{tabular}
	\caption{Hyper-parameter effect on statistics, where the various architectures are labelled based on the number of layers (\emph{ly}), number of cells (\emph{cells}), initial sequence length (\emph{p}) and training-dataset dimension (\emph{N}), expressed in thousands of snapshots. The various model are grouped according to the hyper-parameter on which the analysis focuses, in order: \emph{N}, \emph{cells}, \emph{p} and \emph{ly}. All the modes are employed for  training, accounting for the degeneracy. The validation losses associated with the low- (1) and high-frequencies (2) are also reported. The results in boldface represent the overall best performance for the error over $u_{\rm{RMS}}$ or for the validation losses. A non-physical behaviour arises in the LSTM architectures highlighted in italics. The underlined models show results averaged from three different training. Statistics obtained over $4,000$ time units.}
	\label{LSTM_stats_Hpar}
\end{table}

\begin{figure}[h]
    \centering
    \includegraphics[width=4.6in]{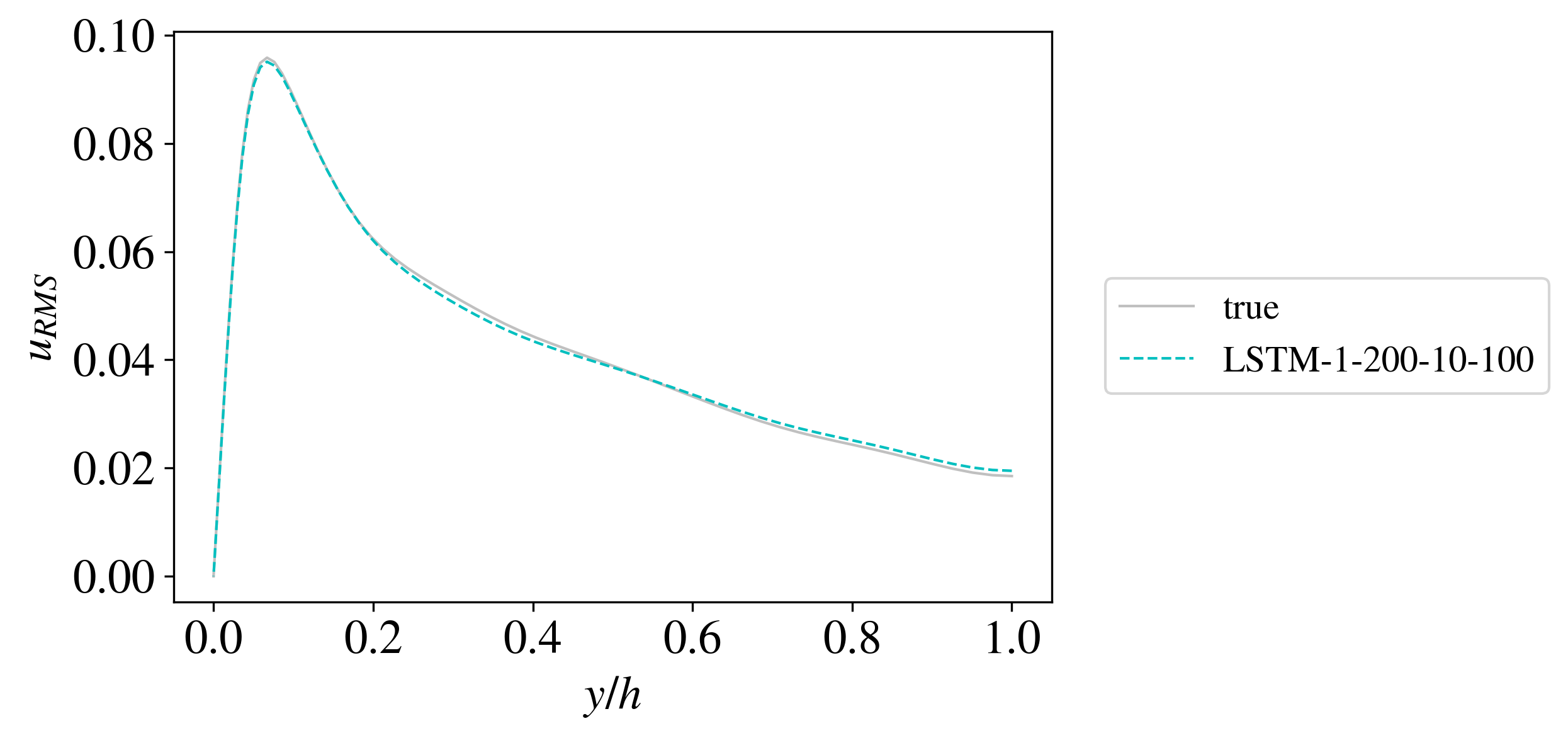}
    \caption{Predicted profiles of the streamwise fluctuations (RMS) for the best model (LSTM-1-200-10-100), compared with the reference. The profiles are averaged with respect to the centerline of the channel ($y=h$).}
    \label{LSTM_stats_pred}
\end{figure}
When dealing with turbulent flows the analysis of the statistical behaviour is a relevant tool to assess the correctness of our predictions. The statistics are derived from the reconstructed velocity fields for a time horizon of $4,000$ time units with a time step of $\Delta t_{\rm{recon}} = 0.8$. We then compare the predicted root-mean-square streamwise velocity-profile, $u_{\rm{RMS}}$, with the one of the original simulation. We define the relative error $E_{u_{\rm{RMS}}}$ for the streamwise fluctuations and report this quantities (together with the validation losses) in \cref{LSTM_stats_Hpar} for all the architectures under study.
The smallest errors for the statistics are obtained for the LSTM-1-200-10-100 architecture. The best model associated with this network generates an error of the fluctuations in the streamwise direction of $E_{u_{\rm{RMS}}} =  1.3 \%$ (note that the results in the table are averaged over 3 networks). The profiles of the true and predicted statistics are illustrated in \cref{LSTM_stats_pred}.
The best architecture for each hyper-parameter analysis is underlined in \cref{LSTM_stats_Hpar} and for these cases three different LSTM networks have been trained with the same setup, for both frequency groups, in order to enforce the consistency of the results (average of the errors is reported), due to the stochastic nature of the training process. \Cref{LSTM_stats_best_stddev} includes the variance associated with the errors on the streamwise velocity fluctuations for these models.
\begin{table}[h]
\centering
	\begin{tabular}{| c || c | c | c |}
	\hline \textbf{model} & LSTM-1-200-10-100 & LSTM-1-150-10-50 & LSTM-1-200-5-50 \\
	\hline 
	$\sigma[E_{u_{\rm{RMS}}}]$ & 1.17 & 0.43 & 0.78\\
	\hline
	\end{tabular}
	\caption{Variance of the errors calculated over the streamwise velocity fluctuations for the best models related to each hyper-parameter analysis, \emph{i.e.} the underlined architectures in table~\ref{LSTM_stats_Hpar}.}
	\label{LSTM_stats_best_stddev}
\end{table}

\subsection{Analysis of the chaotic and dynamic behaviour}
One more useful investigation to assess that the predictions lead to reasonable results and to an appropriate reproduction of the physics in the minimal channel is to study the chaotic behaviour by means of the Lyapunov exponent ($\lambda$). This method allows to verify the sensitivity of a chaotic system to the initial conditions. Given two trajectories we introduce an infinitesimal perturbation $\delta \mathbf{A}_0$ and we observe how the difference between the original and the perturbed trajectories $\delta \mathbf{A}$ evolves in time. We assume the initial divergence to be exponential and to be approximated by:
\begin{equation}
    |\delta \mathbf{A}(t)| \approx e^{\lambda t}|\delta \mathbf{A}_0|,
\end{equation}
where the time evolution of the divergence $|\delta \mathbf{A}(t)|$ is given by:
\begin{equation}
    |\delta \mathbf{A}(t)| = \bigg[\sum_{i=1}^{n} (a_{i,o}(t)-a_{i,p}(t))^2 \bigg]^{1/2},
\end{equation}
where the subscript $p$ represents the perturbed evolution and $o$ the original one. The contribution of all the $n$ modes is summed up together. 

A new dataset of velocity fields is generated after having introduced an initial perturbation of the order of $10^{-3}$ at $t_0 = 10,000$ as a mean energy density of the noise, which translates to a disturbance of $|\delta \mathbf{A}_0| \approx 10^{-7}$ onto the temporal coefficients. The temporal coefficients are derived by projecting the fields on the spatial basis identified in \cref{sec:computational_framework}.
\begin{figure}[h]
    \centering
    \includegraphics[width=2.3in]{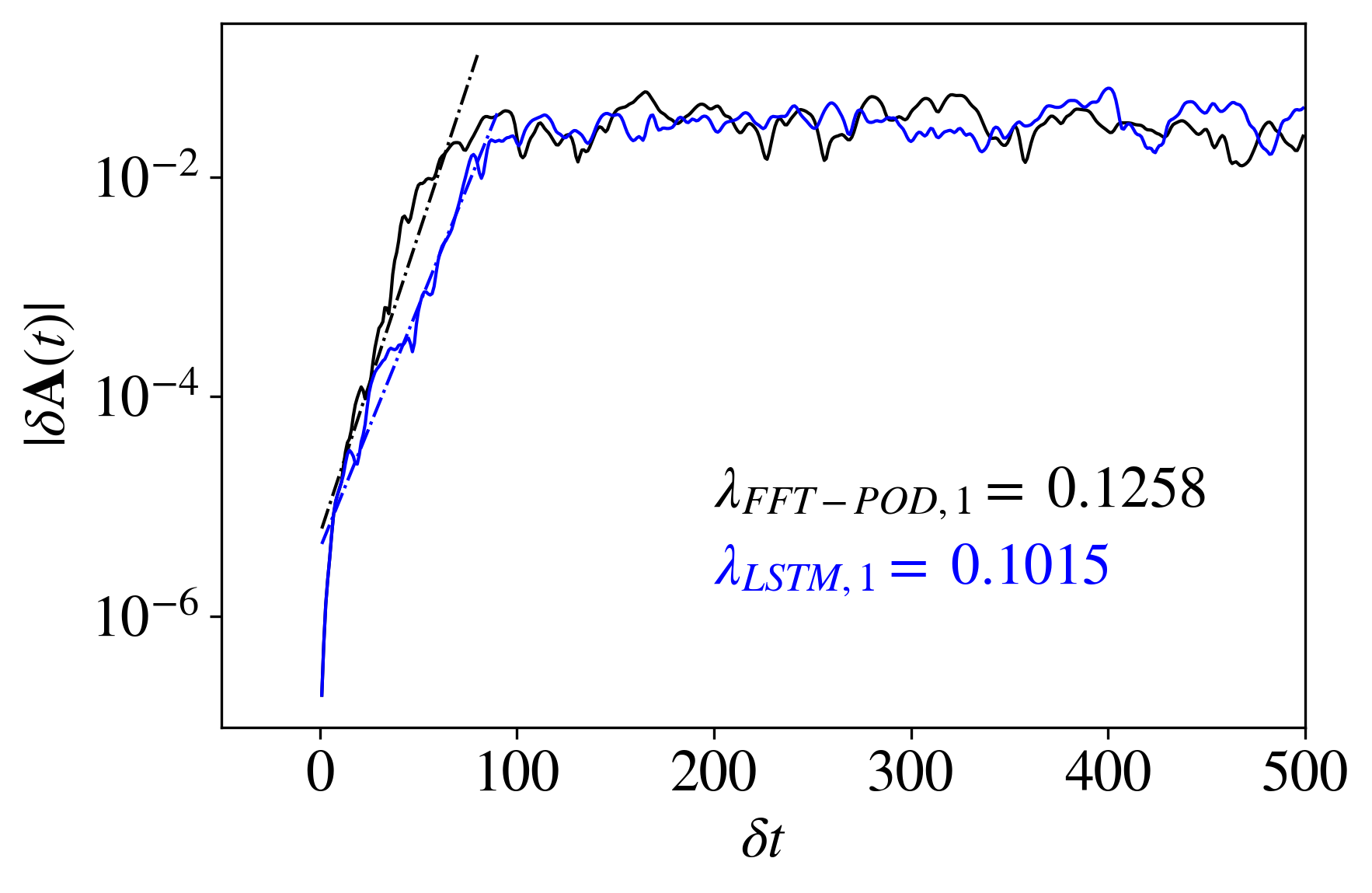}
    \includegraphics[width=2.3in]{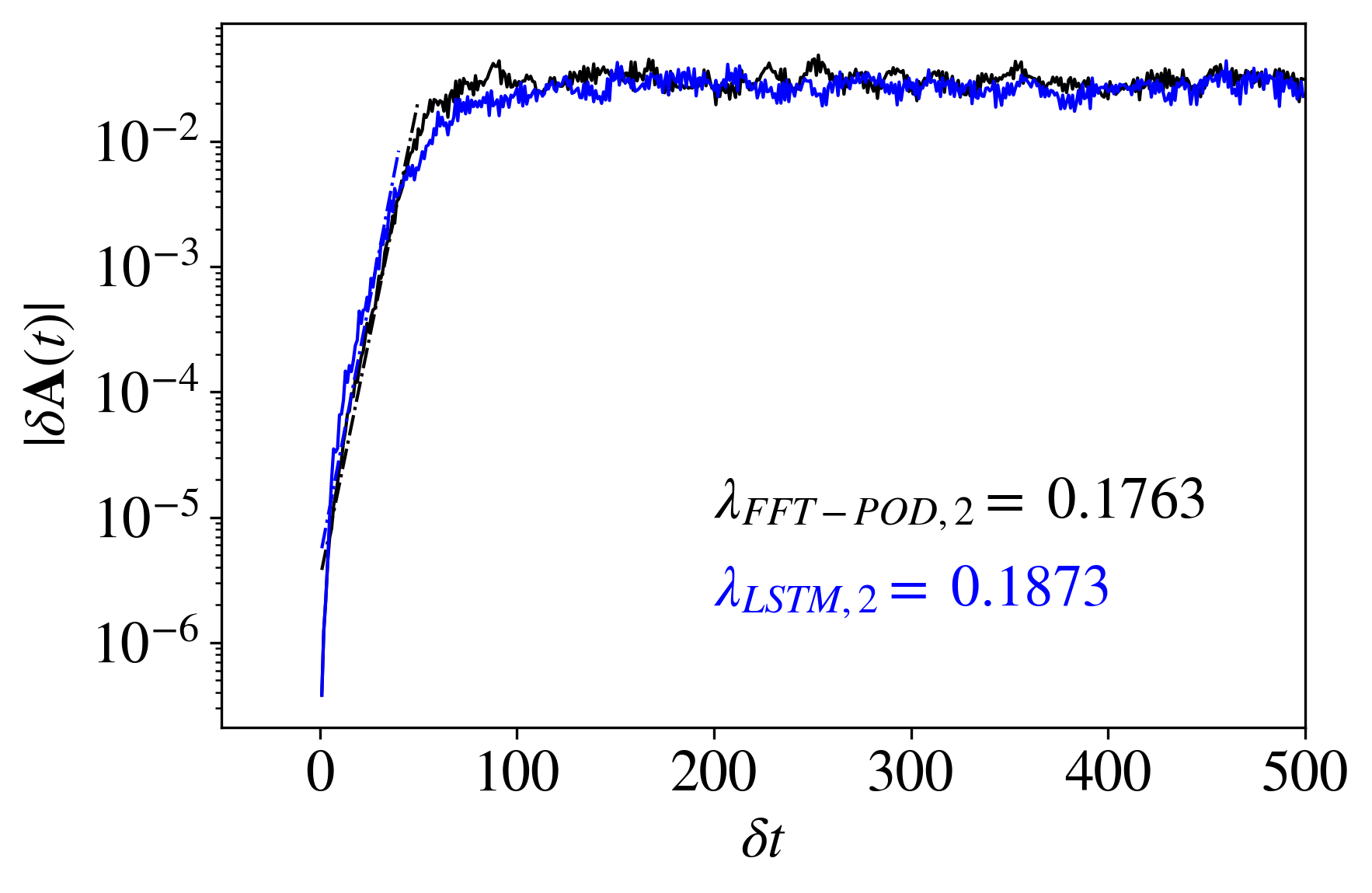}
    \caption{Comparison of the divergence in time between trajectories for the data-driven model (\emph{black}) and for the predicted data from the LSTM-1-200-10-100 architecture (\emph{blue}) once an initial perturbation of $|\delta \mathbf{A}_0| \approx 10^{-7}$ is introduced at $t_0 = 10,000$. Example of Lyapunov exponents for the real part of the low-frequency signals (\emph{left}) and imaginary part of the high-frequency signals (\emph{right}).}
    \label{lyap_picture}
\end{figure}
The analysis is performed separately for the real and imaginary parts of the coefficients as well, as for the low-frequency signals and the high-frequency ones. The predicted perturbed dataset is obtained by employing the LSTM-1-200-10-100 network. \Cref{lyap_picture} shows an example of the evolution in time of the divergence between the trajectories, where the calculated Lyapunov exponents are shown for each case.
\begin{table}[h]
\centering
\begin{tabular}{ p{2cm} }
 \\
 \\
 Reference \\
 \hline
 LSTM \\
\end{tabular}
\begin{tabular}{ |c|c| }
 \hline
 \multicolumn{2}{|c|}{\textbf{real}} \\
 \hline
 $\lambda_1$ & $\lambda_2$ \\
 \hline
 \hline
 0.1258 & 0.1387 \\
 \hline
 0.1015 & 0.0852 \\
 \hline
\end{tabular}
\begin{tabular}{ |c|c| }
 \hline
 \multicolumn{2}{|c|}{\textbf{imaginary}} \\
 \hline
 $\lambda_1$ & $\lambda_2$ \\
 \hline
 \hline
 0.1167 & 0.1763 \\
 \hline
 0.1104 & 0.1873 \\
 \hline
\end{tabular}
\caption{Numerical comparisons of the Lyapunov exponents for the reference model and the LSTM predictions. Note that 1 and 2 denote the low- and the high-frequency models, respectively.}
\label{lyap_compare}
\end{table}
Generally, having $\lambda > 0$ is an indication that the system is chaotic and, furthermore, from \cref{lyap_compare} we can also conclude that the LSTM networks is able to accurately reproduce the physics of the minimal channel, having a really good agreement between the reference and the predicted Lyapunov exponents. We can also observe from \cref{lyap_picture} how the saturation of the curves, \emph{i.e.} the point where the exponential divergence ends, is found approximately after the same time interval.

Moreover, we want to verify the quality of the predicted dynamic behaviour through the use of the Poincaré maps. These maps represent the intersection of two temporal coefficients $a_{\alpha}$ and $a_{\beta}$ with the hyper-plane $a_{\gamma} = 0$, \emph{i.e.} where the coefficient $a_{\gamma}$ changes its sign (${\rm d} a_{\gamma}/{\rm d} t<0$).
\begin{figure}[h]
    \centering
    \includegraphics[width=2.3in]{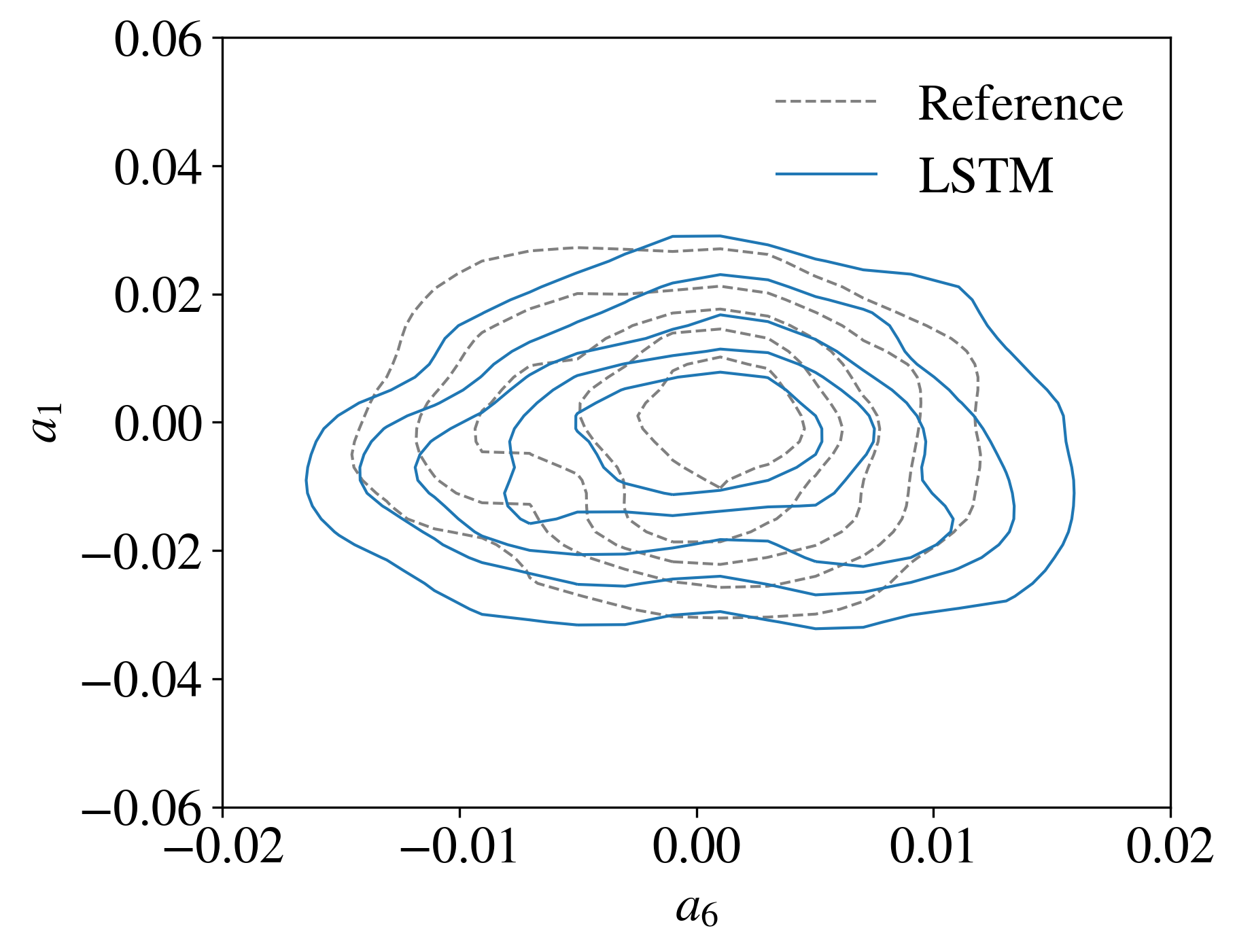}
    \includegraphics[width=2.3in]{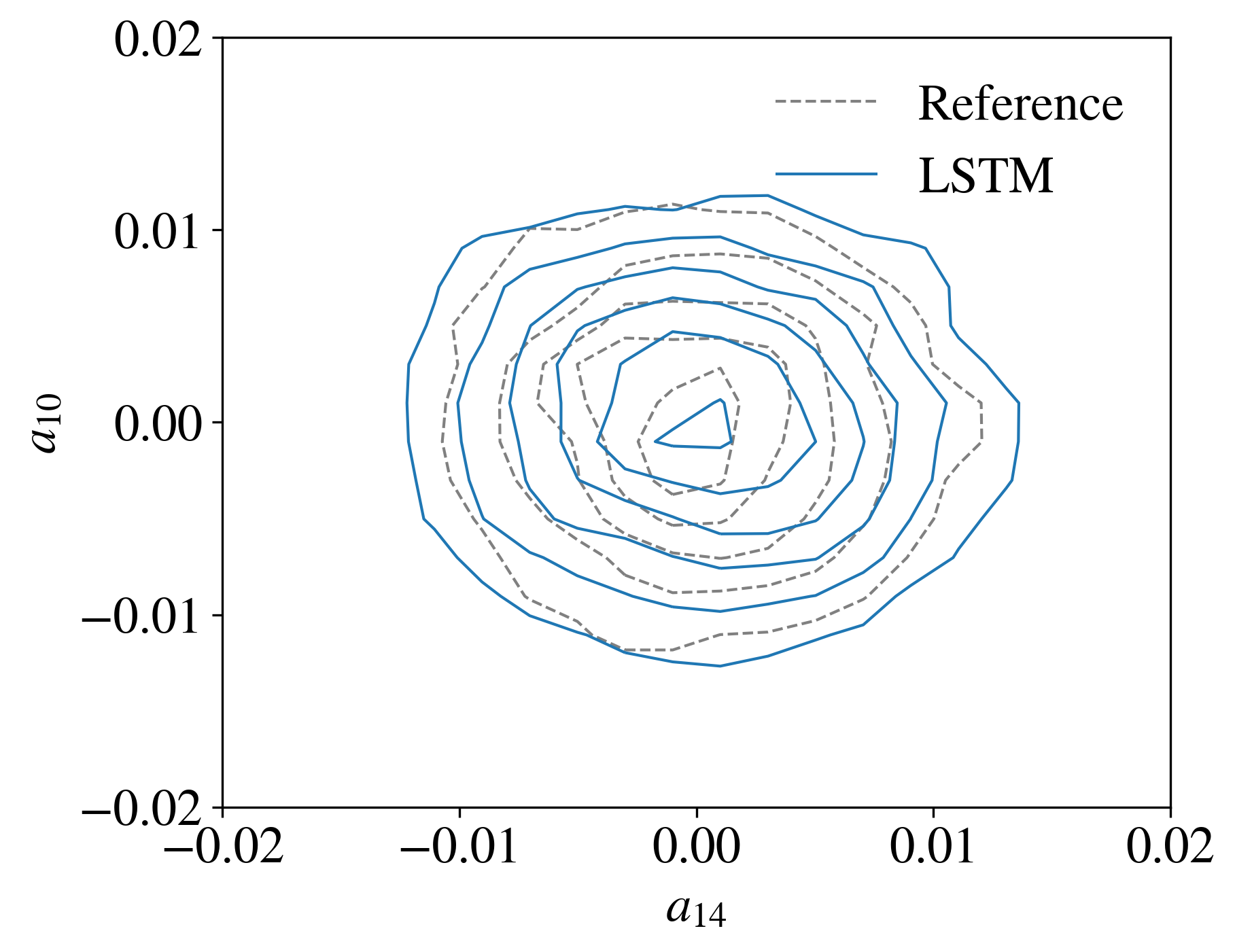}
    \caption{Example of comparison of the Poincaré maps for the most-energetic temporal coefficients belonging to group $\mathbf{[1]}$ (\emph{left} - real coefficients: $a_1$, $a_6$, intersection with plane $a_4 = 0$) and to group $\mathbf{[2]}$ (\emph{right} - imaginary coefficients: $a_{10}$, $a_{14}$, intersection with plane $a_{12} = 0$). The reference maps are reported in grey and the LSTM-predicted ones in blue.}
    \label{poincar_picture}
\end{figure}
Again, this study is performed separately for the real and imaginary part and employs the multi-step concept. The most energetic temporal modes are analysed and reported in \cref{poincar_picture}: intersection of $a_1 - a_6$ with the plane $a_3 = 0$ for group $\mathbf{[1]}$ and $a_{10} - a_{14}$ with $a_{12} = 0$ for group $\mathbf{[2]}$. The picture shows a good agreement, leading to the conclusion that the LSTM network is capable of reproducing the correlation between the amplitudes of the modes which are considered. The same good agreement is observed for the intersection of the other temporal coefficients, associated with a lower energy content, with the hyper-planes identified by the other modes belonging to the data-driven model. Analyzing the results obtained in \cite{Eivazi_koopman} and \cite{srinivasan}, it can be argued that in the data-driven case the reproduction is not as accurate as for the nine-equations model, but this further supports the idea of having a more complex problem to study when dealing with turbulence generated by the minimal channel.

\section{Predictions with Koopman-based framework}
\label{sec:KNF_pred}

An alternative to predict the temporal evolution for a high-dimensional system which is characterized by a non-linear dynamics is to exploit the Koopman-operator theory. The idea is to employ a linear operator on an infinite-dimensional space to describe a non-linear behaviour in a finite domain.

Dynamic-mode decomposition (DMD) offers a first example through which we are able to describe a non-linear problem by using linear observables, i.e. data which is rearranged in a vector-valued snapshot sequence to describe the state-vector of the system~\citep{Schmid_DMD,Tu_DMD}. In this approach, the available linear functions might not be enough to describe the non-linear behaviour of the flow. A solution to this issue is offered by the extended version of the DMD (EDMD) routine~\citep{Williams_EDMD}. Still a limiting factor arises, \emph{i.e.} the necessity of having a previous knowledge of the dynamics of the system in order to define the appropriate dictionary of linear functions. Data-driven approaches have been included in the procedure and are able to provide a rich feature space by means of time-delay embedding instead of having to define linear/non-linear observable functions~\citep{Li_EDMD_datadriven,Lusch2018,Takeishi_DMD_Koopman_invariant}. Following this idea, delay embedding has been integrated with DMD in the Hankel-DMD method (HDMD) to study chaotic systems~\citep{arbabi_mezic}. In the study from \cite{brunton_chaos} this model was extended by proposing the Hankel alternative view of Koopman (HAVOK), which incorporates Koopman theory, time-delay embedding and sparse regression. Moreover, \cite{khodkar_koopman} presented a new Koopman-based framework in which nonlinearities are modelled through an external forcing and it is capable of successfully predicting the dynamics of highly-chaotic systems.

\subsection{Koopman with non-linear forcing (KNF)}
The preference of the Koopman framework which employs a non-linear forcing (KNF) over the HDMD is driven by the better accuracy achieved by the first method in the predictions of a chaotic dynamical system~\citep{khodkar_koopman}. We consider a dynamical system:
\begin{equation}
    \mathbf{x}_{t+1} = \mathbf{F}(\mathbf{x}_t),
    \label{dyn_sist}
\end{equation}
on the state space $\mathcal{M}\subseteq \mathbb{R}^{m'}$ , where $\mathbf{x}$ is a spatial coordinate of the state, and $\mathbf{F}:\mathcal{M} \to \mathcal{M}$ is the operator describing the evolution of the system. The Koopman operator $\mathbfcal{K}$ acts on functions of state space (called \emph{observables}) $g:\mathcal{M}\to\mathbb{C}$ as follows:
\begin{equation}
    \mathbfcal{K} g = g \circ\mathbf{F},
\end{equation}
where $\circ$ denotes the composition of $g$ with $\mathbf{F}$. The Koopman operator defines a new dynamic of the system which governs the evolution of the \textit{observables} $g_{t} = g(\mathbf{x}_t)$ in discrete time and which is linear and infinite-dimensional.
We then move to the Koopman-based framework:
\begin{equation}
    \mathbf{x}^{t+1} = \mathbf{A}\mathbf{x}^t + \mathbf{B} \mathbf{f}^t,
    \label{dyn_syst_forcing}
\end{equation}
where the external forcing $\mathbf{f}$ models the non-linearities and $\mathbf{A}$ and $\mathbf{B}$ denote the unknown matrices which describe the dynamical system. By knowing these matrices we are able to advance the solution of the state in time.
The vector $\mathbf{f}$ includes any candidate nonlinear functions of $\mathbf{x}$, described by polynomials:
\begin{equation}
    \mathbf{f}^{i} = \begin{bmatrix}
        1& (\mathbf{x}^{i})^{p_{2}}& (\mathbf{x}^{i})^{p_{3}} & \cdots & (\mathbf{x}^{i})^{p_{n}}
    \end{bmatrix}^{\mathsf{T}},
\end{equation}
where $\mathsf{T}$ denotes the transpose of the matrix. Here, for instance, $(\mathbf{x}^{i})^{p_{2}}$ and $(\mathbf{x}^{i})^{p_{3}}$ indicate any possible quadratic or cubic non-linearities, respectively ($a^{i}_{j}a^{i}_{k}$ and $a^{i}_{j}a^{i}_{k}a^{i}_{l}$, where $j,k,l$ denote the modes which are considered). Constant or sinusoidal functions can also be considered to build this vector.
Usually, an intuition of the governing equations of our problem is necessary in order to define the forcing term but, on the other hand, the sparse identification of non-linear dynamics (SINDy) method~\citep{brunton_sindy} offers an alternative to achieve this in an optimized way, thanks to the sparsity of the matrix which is considered.
We first have an iterative linear regression of $\begin{bmatrix}
    \mathbf{x}^{2} & \mathbf{x}^{3} & \cdots & \mathbf{x}^{N}
\end{bmatrix}$ on $\begin{bmatrix}
    \mathbf{xf}^{1} & \mathbf{xf}^{2} & \cdots & \mathbf{xf}^{N-1}
\end{bmatrix}$, where $ \mathbf{xf}^{i} = \begin{bmatrix}
    \mathbf{x}^{i} & \mathbf{f}^{i}
\end{bmatrix}^{\mathsf{T}}$, and then zero out all the coefficients which are smaller than a threshold value $\varepsilon$. This procedure is performed in an iterative way until a convergence is observed for the non-zero coefficients. \Cref{algo:SINDy} shows a schematic of the SINDy method, which has an important advantage over deep-learning methods due to its interpretability. Note that this issues of interpretability of deep-learning models are discussed by \cite{vinuesa_interpretability}.

\SetKwComment{Comment}{$\triangleright$\ }{}
\begin{algorithm}[h]
    \SetKw{KwInit}{Initialize}
    \DontPrintSemicolon
    \KwIn{$\mat{y} = \begin{bmatrix}
        \mathbf{x}^{2} & \mathbf{x}^{3} & \cdots & \mathbf{x}^{N}
    \end{bmatrix},~\mat{x} = \begin{bmatrix}
        \mathbf{xf}^{1} & \mathbf{xf}^{2} & \cdots & \mathbf{xf}^{N-1}
    \end{bmatrix}$, threshold $\varepsilon$}
    \KwOut{$\mathbf{I}_{\mathrm{active}}$ \Comment*[r]{Indices of active nonlinearities}}
    \vspace{0.05in}
    $n, m \leftarrow$ number of rows of $\mat{y}, \mat{x}$\;
    \KwInit{$\mathbf{C}(n, m)$ \Comment*[r]{Coefficients}}
    \KwInit{$\mathbf{I}(n, m)$ \Comment*[r]{Active indices, dtype = bool}}
    \vspace{0.05in}
     \For{$i\leftarrow 1$ \KwTo $\mathrm{Max~Iteration = 20}$}{
         \For{$j\leftarrow 1$ \KwTo $n$}{
             $\mathbf{I}_j \leftarrow \mathbf{I}[j, ~:]$\;
            $\mathbf{C}[j,~\mathbf{I}_j] \leftarrow \mathrm{Ridge~Regression}(\mat{y}[j,~:], \mat{x}[\mathbf{I}_j ,~:])$\;
            $\mathbf{I}_j \leftarrow \mathrm{abs}(\mathbf{C}[j,~:]) >= \varepsilon$ \Comment*[r]{Find big coefficients}
            $\mathbf{C}[j,~\sim\mathbf{I}_j] \leftarrow 0$ \Comment*[r]{Zero out small coefficients}
            $\mathbf{I}[j,~:]  \leftarrow \mathbf{I}_j$
         }
        \If{$\mathbf{C}$ \rm{does not change}}{
            \textbf{Break}\;
        }
     }
    $\mathbf{I}_{\mathrm{active}} \leftarrow$ \rm{Maximum element of each column in }$\mathbf{I}[:,~n:]$\;
     \caption{SINDy algorithm to compute non-linear terms for the KNF method.}
     \label{algo:SINDy}
\end{algorithm}

At this point we employ the Hankel-matrix representation of the data to redistribute the values in a matrix form, so that we express the state vector and the forcing term as $\mathbf{X}$ and $\pmb{\mathcal{F}}$, respectively:
\begin{equation}
    \mathbf{X}=\begin{bmatrix} 
    \mathbf{x}^1 & \dots  & \mathbf{x}^{N'-q}\\
    \vdots  & \ddots & \vdots \\
    \mathbf{x}^q & \dots  & \mathbf{x}^{N'-1}
    \end{bmatrix},
    \quad
    \pmb{\mathcal{F}}=\begin{bmatrix} 
    \mathbf{f}^1 & \dots  & \mathbf{f}^{N'-q}\\
    \vdots & \ddots & \vdots \\
    \mathbf{f}^q & \dots  & \mathbf{f}^{N'-1}
    \end{bmatrix},
    \label{hankel_mat}
\end{equation} 
and equation \eqref{dyn_syst_forcing} is expressed in its time-embedded form as:
\begin{equation}
    \mathbf{X}^{t+1} = \mathbf{A}\mathbf{X}^t + \mathbf{B} \pmb{\mathcal{F}}^t.
    \label{dyn_syst_forcing_emb}
\end{equation}
The matrices $\mathbf{X}$ and $\pmb{\mathcal{F}}$ have sizes $(m' \times q) \times (N'-q+1)$ and $(n' \times q) \times (N'-q)$ respectively, where $m'$ is the number of state variables, $n'$ is the size of the forcing vector (usually $n' >> m'$), $N'$ is the number of vector-valued observables and $q$ is the delay-embedding dimension. We rearrange the data in the Hankel matrices following the exact-DMD algorithm formulation~\citep{arbabi_mezic}, and we define $\pmb{X}'$ and $\pmb{Y}'$ as:
\begin{equation}
    \pmb{X}' = \begin{bmatrix} \pmb{\mathcal{X}}^1 \cdots \pmb{\mathcal{X}}^{N'-q} \end{bmatrix}, \quad \pmb{Y}' = \begin{bmatrix} \pmb{\mathcal{X}}^2 \cdots \pmb{\mathcal{X}}^{N'-q+1} \end{bmatrix},
    \label{Hankel_arrange}
\end{equation}
where $\pmb{\mathcal{X}}^i$ denotes the $i^{th}$ column of the Hankel matrix. Ultimately, the matrices \textbf{A} and \textbf{B} are derived by using the DMDc algorithm ($c$ stands for control) introduced by \cite{proctor_mode_decomp} which relies on the minimization of the Frobenius norm $||\pmb{Y}'- \mathbf{A}\pmb{X}' - \mathbf{B} \pmb{\mathcal{F}}||_F$. These are derived as:
\begin{equation}
   \mathbf{A} = \Hat{\mathbf{U}}^{\mathsf{H}}\pmb{Y}'\Tilde{\mathbf{V}}\Tilde{\mathbf{S}}^{-1} \Tilde{\mathbf{U}}_1^{\mathsf{H}}\Hat{\mathbf{U}}, \quad \quad \quad \quad 
   \mathbf{B} = \Hat{\mathbf{U}}^{\mathsf{H}}\pmb{Y}'\Tilde{\mathbf{V}}\Tilde{\mathbf{S}}^{-1}\Tilde{\mathbf{U}}_2^{\mathsf{H}}.
   \label{KNFmapping}
\end{equation}
The energy truncation of the matrices derived from the SVD, on $\pmb{Y}'$ and $\big[ \pmb{X}' \quad \pmb{\mathcal{F}}\big]$, is denoted by $\Hat{(\cdot)}$ and $\Tilde{(\cdot)}$, respectively. The truncation rank of $\pmb{Y}'$ is $r$ and it leads to the decomposition $\pmb{Y}' = \Hat{\mathbf{U}} \Hat{\mathbf{S}} \Hat{\mathbf{V}}^{\mathsf{H}}$, with 
$\Hat{\mathbf{U}} \in \mathrm{I\!R}^{(m' \times q) \times r}$, $\Hat{\mathbf{S}} \in \mathrm{I\!R}^{r \times r}$ and $\Hat{\mathbf{V}} \in \mathrm{I\!R}^{(N'-q) \times r}$. In the same way $\big[ \pmb{X}' \quad \pmb{\mathcal{F}}\big] = \Tilde{\mathbf{U}} \Tilde{\mathbf{S}} \Tilde{\mathbf{V}}^{\mathsf{H}}$. The truncation rank is defined by $d$ for this case, with $\Tilde{\mathbf{U}} \in \mathrm{I\!R}^{((m'+n')\times q) \times d}$, $\Tilde{\mathbf{S}} \in \mathrm{I\!R}^{d \times d}$ and $\Tilde{\mathbf{V}} \in \mathrm{I\!R}^{(N'-q) \times d}$. The choice of the truncation defines a threshold below which the contribution of the modes is considered as negligible from the energetic perspective and is based on SVD rank-truncation methods such as the optimal hard threshold illustrated by \cite{gavish_optimal_hard_threshold}.
Finally, $\Tilde{\mathbf{U}}$ is divided into $\Tilde{\mathbf{U}} = \big[ \Tilde{\mathbf{U}}_1^{\mathsf{H}} \quad \Tilde{\mathbf{U}}_2^{\mathsf{H}} \big]$, where $\Tilde{\mathbf{U}}_1 \in \mathrm{I\!R}^{(m' \times q) \times d}$ enters the computation for $\mathbf{A}$ and $\Tilde{\mathbf{U}}_2 \in \mathrm{I\!R}^{(n' \times q) \times d}$ defines $\mathbf{B}$, which is related to the forcing.

\subsection{KNF temporal predictions}
As for the LSTM case, the multi-step concept has been applied. This time the training consists of finding the matrices \textbf{A} and \textbf{B}, underlining the non-stochastic nature of this process, that lead to have always the same model for the same setup of parameters on which predictions are based (LSTM is deterministic only once it is trained).
For all the tests performed with different configurations the method is not capable of capturing the physics of the system, as discussed below.

\begin{figure}[h]
    \centering
    \includegraphics[width=3.3in]{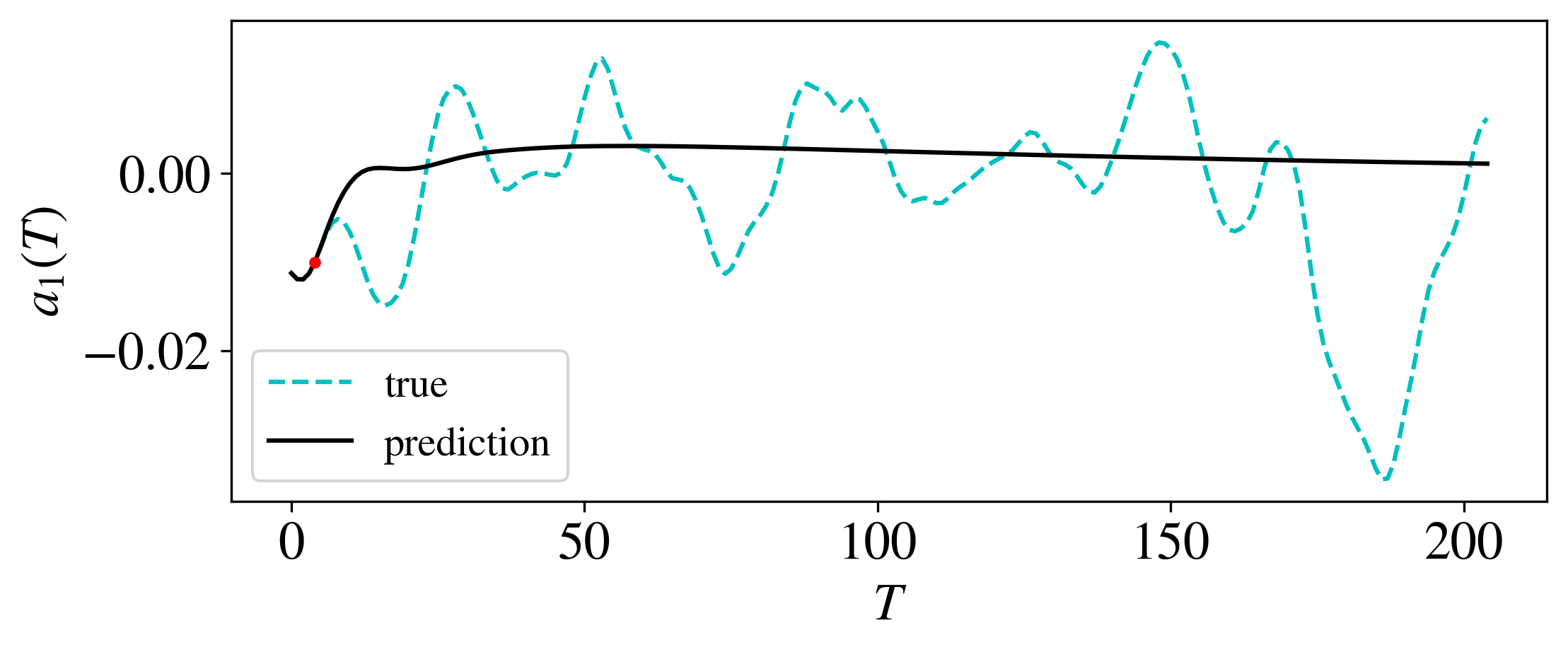}
    \includegraphics[width=1.42in]{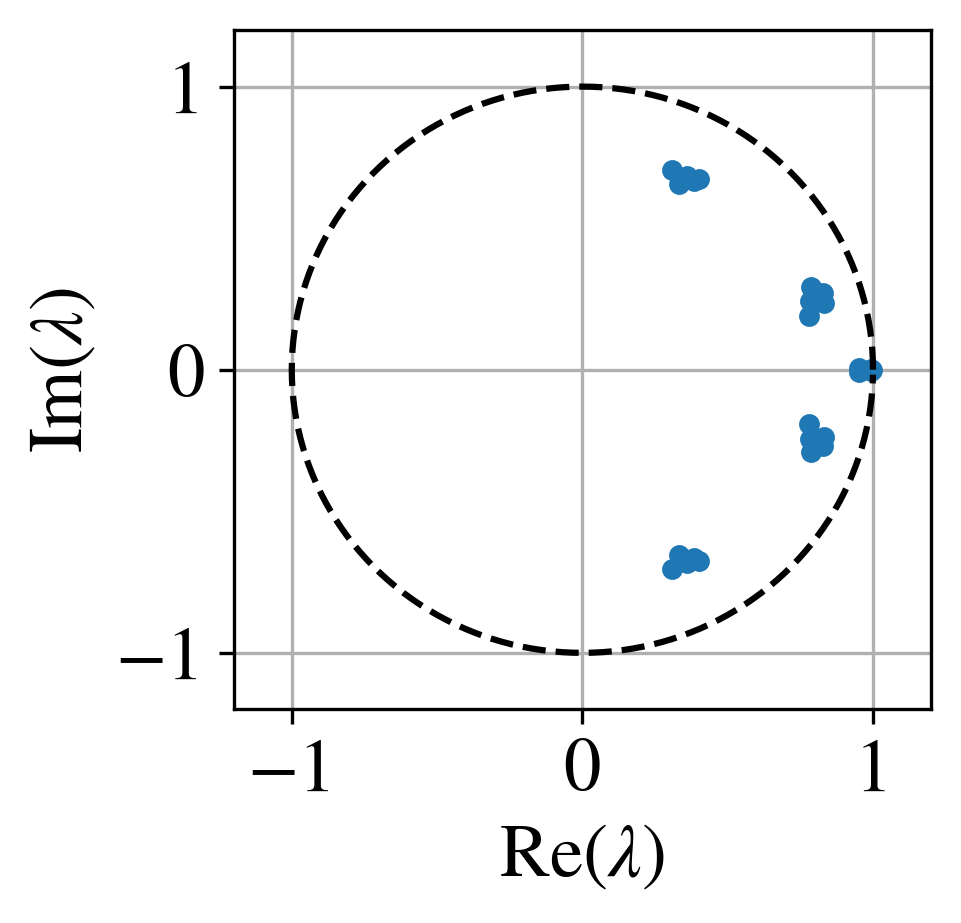}
    \caption{Short-term prediction of the $a_1$ coefficient with $q = 5$ on the left. The unitary circle representing the eigenvalues of matrix $\mathbf{A}$ is shown on the right.}
    \label{KNF_a1_pred}
\end{figure}

By assessing the predictions related to the low-frequency group in \cref{KNF_a1_pred} we observe how the predicted fluctuations associated with the temporal modes are not sustained. 
For this case a delay-embedding dimension of $q=5$ was chosen. The dataset comprises over $10,000$ snapshots of the 5 most energetic modes of group $\pmb{[1]}$ with a sampling time $\Delta t_{[1]} = 0.8$. Non-linearities are modelled with a third-order polynomial without employing sparsity promotion. The energy-based tolerances for matrices $\mathbf{A}$ and $\mathbf{B}$ are $e_r = 10^{-5}$ and $e_p = 10^{-5}$. The unitary circle where the eigenvalues of matrix $\mathbf{A}$ are plotted might be considered as an indicator of why the predicted evolution is dampened in time, as most of the eigenvalues are inside this circle ($|\lambda| < 1$), thus a decay is expected. Different tests have been performed by adjusting the tolerances $e_r$ and $e_p$ or by either gradually increasing the dataset dimension or employing more modes, without showing relevant improvements. The modeling of the forcing term has been attempted with different polynomial orders and also by using trigonometric functions. Note that sparsity promotion did not produce noteworthy changes either. 
Finally, as $q$ increases the fluctuations are sustained for a longer time horizon, but eventually they are flattened out too. One option would be to further increase the delay-embedding dimension, but this requires more data in the initial sequence leading to predictions which are not efficient.

The same method has been applied also to the high-frequency modes, showing a similar trend. The sampling time is of $\Delta t_{[2]} = 0.2$ and non-linearities are modelled with a second-order polynomial. Tolerances are the same as for the low-frequency case and sparsity is promoted. The dataset dimension consists of $10,000$ snapshots and it still considers the 5 most energetic modes for this group.
\begin{figure}[h]
    \centering
    \includegraphics[width=3.3in]{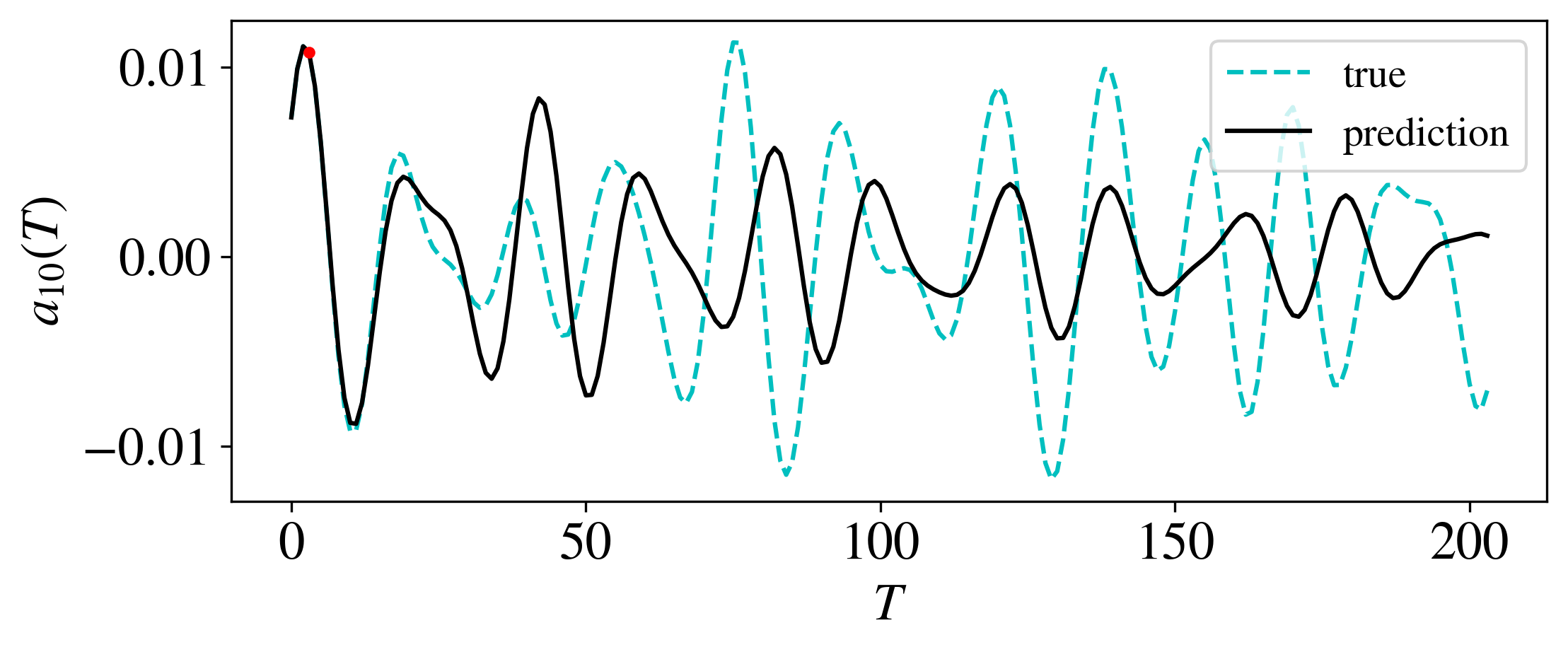}
    \includegraphics[width=1.42in]{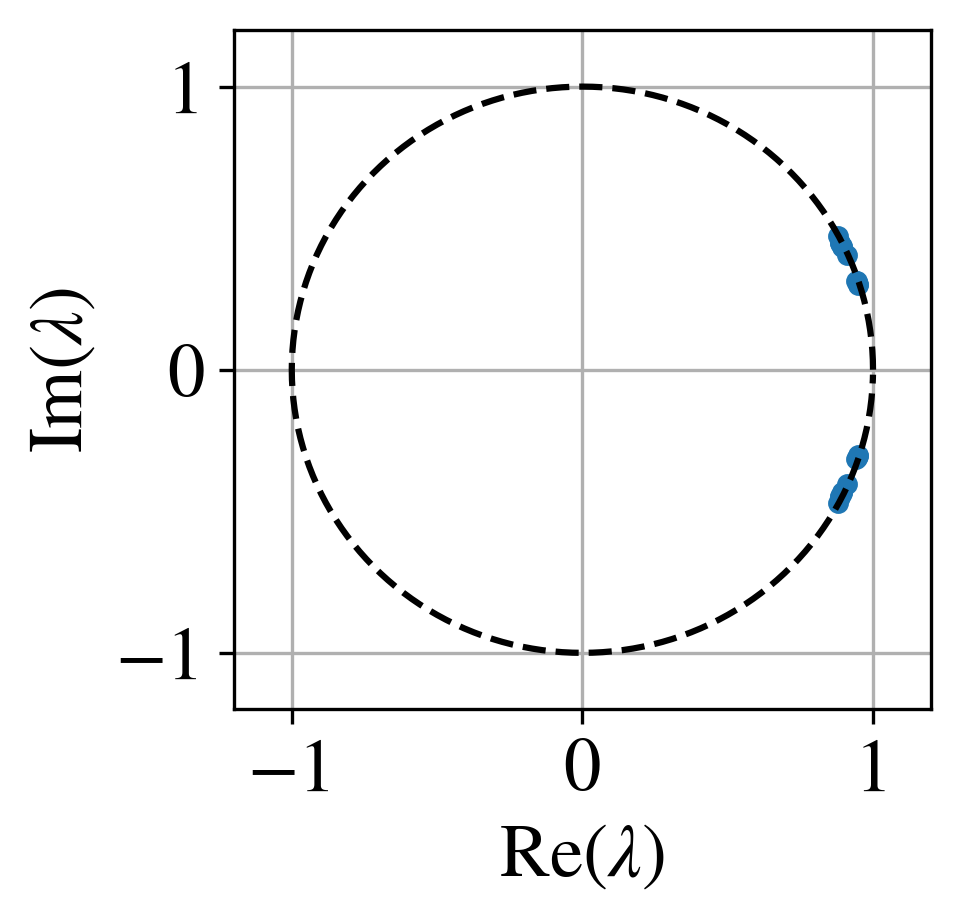}
    \caption{Short-term prediction of the $a_{10}$ coefficient with $q = 4$ on the left. The unitary circle representing the eigenvalues for the matrix $\mathbf{A}$ is shown on the right.}
    \label{KNF_a10_pred}
\end{figure}
The delay-embedding dimension is $q=4$ for the example reported in \cref{KNF_a10_pred} and this time the opposite behaviour is observed, \emph{i.e.} as we increase $q$ the fluctuations are dampened out earlier, whilst they are sustained for a wider time span for the low-frequency signals.
For both the low- and high-frequency signals the forcing term has been modelled with polynomials up to the 7-$th$ order. No significant improvements were observed for the first group of signals, whereas the behaviour of the predicted fluctuations showed some differences for the high-frequency signals. \Cref{KNF_a10_pred_poly6} shows the long-term behaviour of these fluctuations with non-linearities modelled with a 6-$th$ order polynomial and the energy-based tolerances which are $e_r = 10^{-10}$ and $e_p = 10^{-10}$.
\begin{figure}[h]
    \centering
    \includegraphics[width=3.3in]{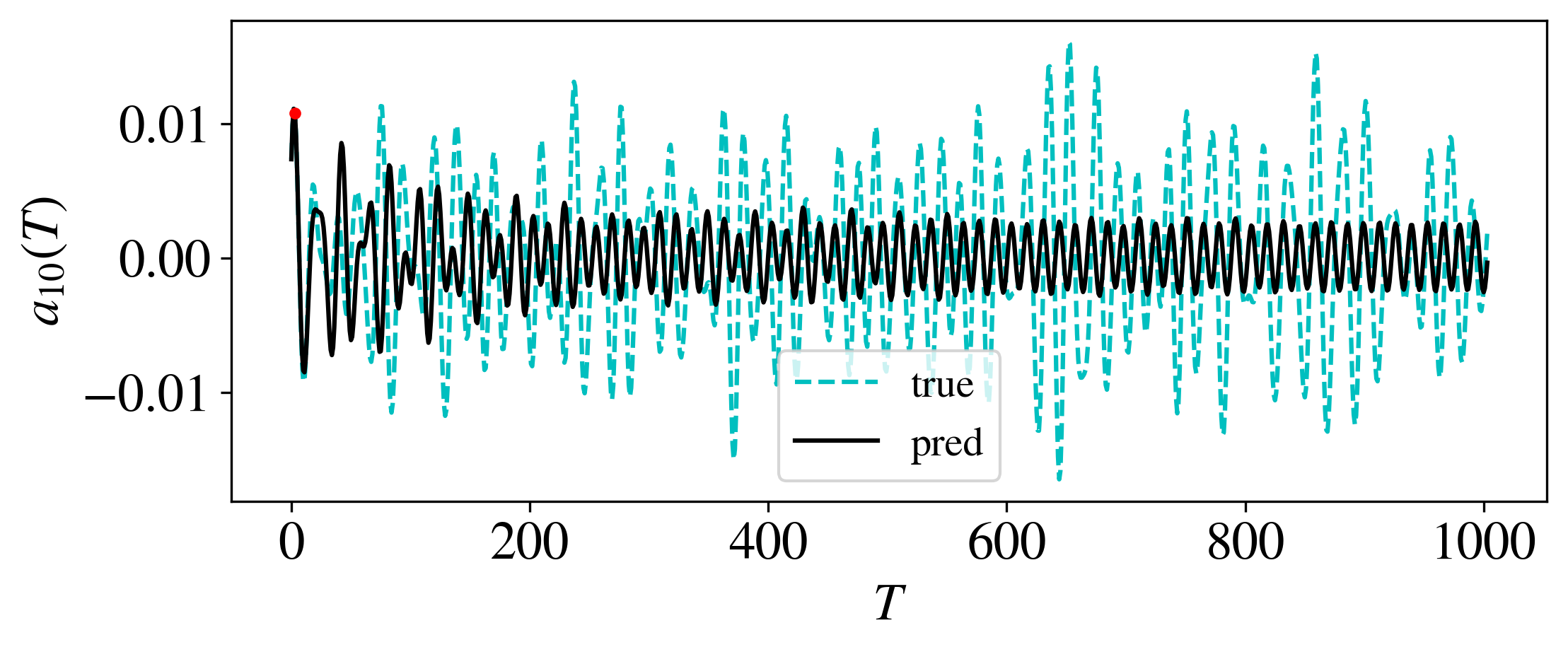}
    \includegraphics[width=3.3in]{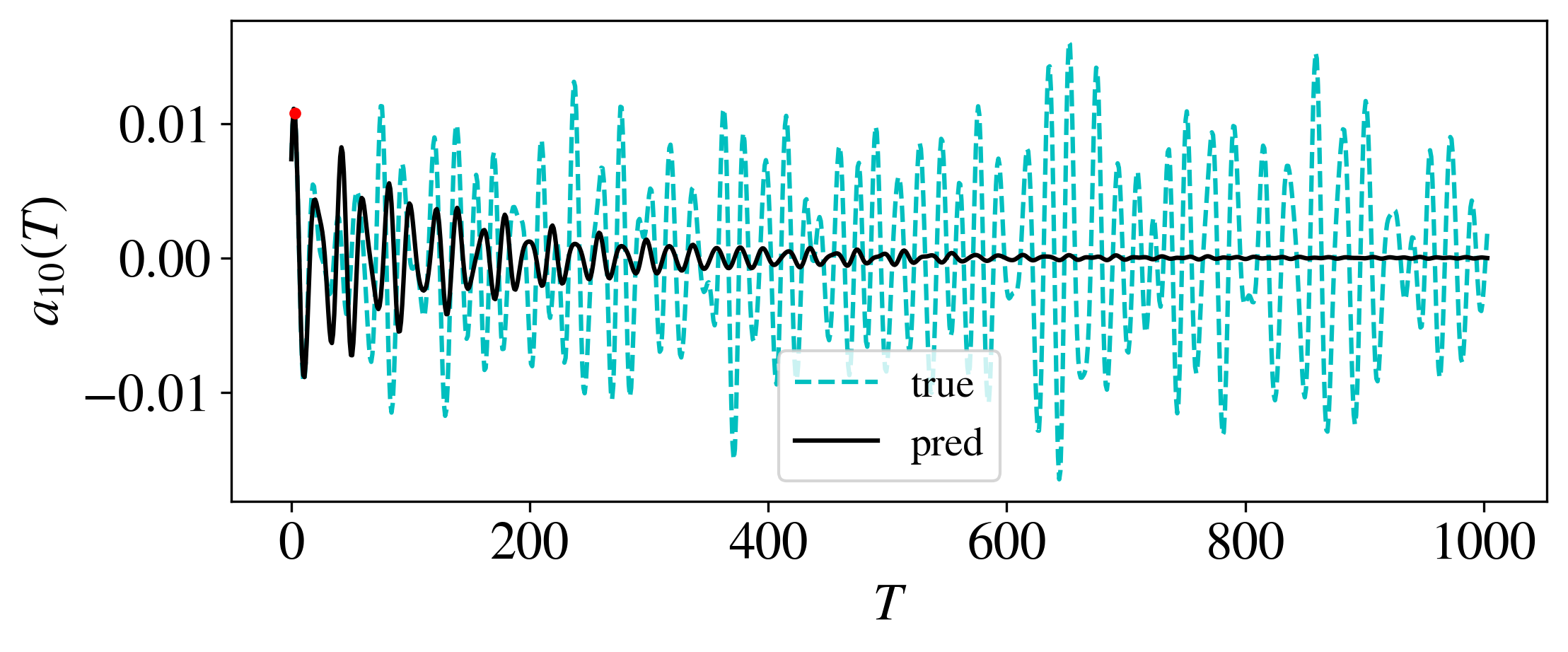}
    \caption{Long-term prediction of the $a_{10}$ coefficient with $q = 4$ and non-linearities modelled with a 6-$th$ order polynomial (\emph{top}) compared with the long-term prediction for the case in \cref{KNF_a10_pred} (\emph{below}).}
    \label{KNF_a10_pred_poly6}
\end{figure}
Although the fluctuations of the predicted signal are now sustained for a longer time horizon, the coefficients are still not in the correct range.

In many tests the eigenvalues of matrix $\mathbf{A}$ are on the unitary circle, as it is reported for the example in \cref{KNF_a10_pred}, but a stable behaviour is not observed. Indeed, if we recall equation \eqref{dyn_syst_forcing}, the system dynamics is not only described by this matrix but also by matrix $\mathbf{B}$, which expresses the contribution of the forcing term to the evolution of the state in time. By extending the analysis in this direction some more information might be available to understand why the method has not led to successful predictions. The analysis of the $\mathbf{B}$ matrix could also highlight the effect of enhancing or not the sparsity when defining the forcing vector $\mathbf{f}$, a study that will be carried out in future work.

\section{Reduced-order models (ROMs)}
\label{sec:ROMs}

In this section we investigate the possibility of implementing reduced-order models (ROMs) and we assess the predictions of an LSTM network on them. The big advantage of these models is given by the possibility to represent turbulence only by using the most relevant structures and, consequently, to handle efficiently the governing physics from large datasets. We also seek a reduction of the time needed to train the network, from the computational point of view. We introduce two different alternatives which are based on using fewer modes for the training of the network with the only difference that $(a)$ in the first case we still want to predict all the 100 modes of the data-driven model (\cref{sec:ROMs_a}) and $(b)$ in the second case we only want to predict the modes which are employed for the training (\cref{sec:ROMs_b}), leading to a reconstruction of the velocity fields which is further simplified, but still able to include the most relevant coherent structures.

The new trained models are built on the identification of sub-groups which share the same features in the homogeneous directions. Six sub-groups are recognized over the first 100 modes, if we exclude the net-flux modes, and have the same wave-number pair. \Cref{LSTM_RoMs_example} shows how the modes are divided and it reports in red an example of the modes which are used for training the new networks. In the example the modes employed for training of what we define ROM1 are highlighted in bold; in this ROM only one mode from each sub-group enters the training process.
\begin{table}[h]
\centering
    \begin{tabular}{ |c||c|c| } 
    \hline
    \multirow{2}{*}{\textbf{Freq. group}} & \textbf{Wavenumber} & \multirow{2}{*}{\textbf{Mode index}} \\
    & \textbf{pair} & \\
    \hline
    \hline
    \multirow{2}{4em}{Low-freq} & $\mathbf{(0,1)}$ &  \textcolor{red}{1}, 4, 7, 9, 24, 26, 42, 60, 64, 94 \\ \cline{2-3}
    & $\mathbf{(0,2)}$ & \textcolor{red}{44}, 46 \\ \cline{2-3}
    \hline
    \hline
    \multirow{5}{4em}{High-freq} & $\mathbf{(1,0)}$ & \textcolor{red}{19}, 21, 37, 39, 47, 50, 63, 74, 95 \\ \cline{2-3}
    & $\mathbf{(1,1)}$ & \textcolor{red}{10}, 12, 14, 16, 29, 30, 32, 34, 51, 54\\
    & &  56, 58, 77, 85, 90, 91\\ \cline{2-3}
    & $\mathbf{(1,2)}$ & \textcolor{red}{78}, 80, 82, 86 \\ \cline{2-3}
    & $\mathbf{(2,1)}$ & \textcolor{red}{67}, 69, 70, 72, 98 \\ \cline{2-3}
    \hline
    \end{tabular}
    \caption{Sub-division of the modes according to the various turbulent features. The modes in red are the ones employed to train the reduced-order model ROM1 (and are also the predicted ones for the ROMs which employ an alternative field reconstruction). Degeneracy is retained back when reconstructing the fields.}
	\label{LSTM_RoMs_example}
\end{table}
A correlation analysis is employed again, this time to understand if the use of fewer modes for a given sub-group is representative of the others.
If the correlation factor between signals of the same group is higher than the correlation between signals belonging to different groups, then the signals of the same sub-group have a similar behaviour and they describe the same turbulent structure. This is observed overall in our case, \emph{i.e.} the correlation factor between signals of the same sub-group is of the order of $10^{-2}$ and it is at least an order of magnitude larger with respect to the correlation factor associated with signals of different sub-groups. This justifies the division in \cref{LSTM_RoMs_example}, thus our networks will be able to capture the dynamics of the sub-group with a reduced amount of information required.

\subsection{Energy-based reconstruction with 100 modes}
\label{sec:ROMs_a}
The first approach still aims to reconstruct the field with 100 modes in order to compare the predictions of the statistics with the ones of the data-driven case (best model in \cref{LSTM_stats_pred}). The difference with respect to the latter case is related to the training, where fewer modes are employed. We name the models ROM1 and ROM2, which represent the models which employ either one or two modes for each turbulent feature, respectively. Following the procedure which is exemplified in \cref{LSTM_RoMs_example}, this means that we are using either 6 or 12 modes out of 100 for the training instead of 46. 
Degeneracy needs to be considered also in this case since, in practice, we only predict one mode for each couple of complex-conjugate modes (thus 46 out of 100, excluding the 7 modes related to the first wavenumber pair).

\begin{figure}[h]
    \centering
    \includegraphics[width=4.6in]{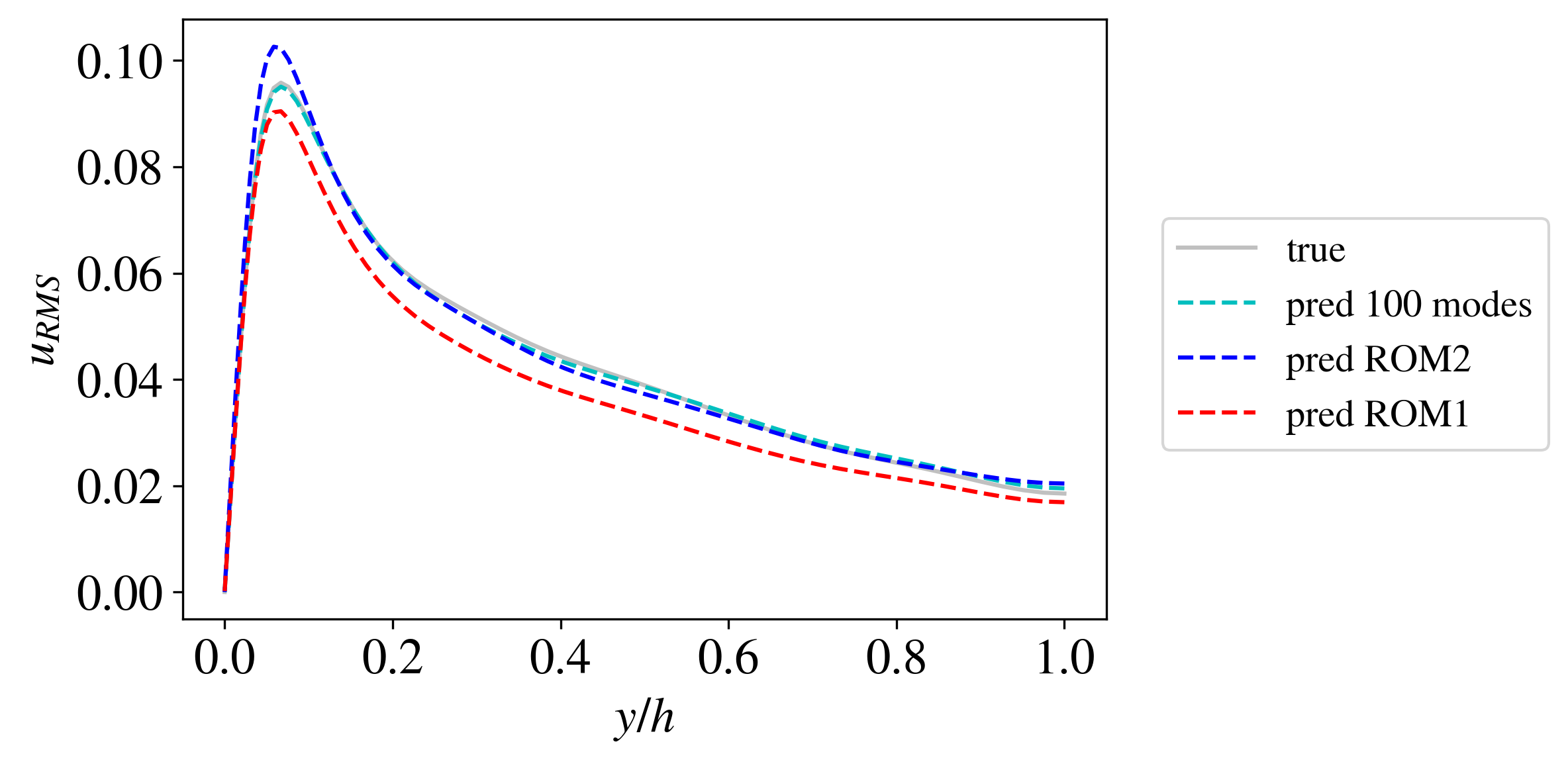}
    \caption{Streamwise velocity fluctuations for the different reduced-order models discussed above. The true profile is reported in grey.}
    \label{ROMa_urms_profiles}
\end{figure}

For the predictions of the reduced models we still employ an LSTM-1-200-10-100 architecture in order to obtain results consistent with the ones reported above. A first aspect which can be noticed is that we achieve a negligible reduction of the time needed to train the network. Interestingly, we can observe a good agreement of the predicted statistics in \cref{ROMa_urms_profiles}, even if the networks employ less information. This is an important benefit when considering a data-driven framework because it shows how turbulence represented by large datasets can be handled efficiently with a reduced amount of data.
In order to achieve statistical significance, we perform 3 separate trainings (which adds up to six due to the multi-step approach) with the same settings of the hyper-parameters for each reduced-order model case. The profiles in \cref{ROMa_urms_profiles} represent the predictions of the best model among the three available ones where the errors on the streamwise fluctuations are $6.1 \%$ for ROM2 and $9.5 \%$ for ROM1. 
Additionally, it can be observed that an overshoot is present in the case where 12 modes are employed for the training, while the opposite happens when only 6 modes define the training dataset. It might be stated that the specific choice of the modes employed in the training process significantly affects the fluctuations. 
\begin{table} [h]
\centering
	\begin{tabular}{| c | c c c |}
	\hline Model  & $E_{u_{\rm{RMS}}}[\%]$ & val. loss 1 & val. loss 2 \\
	\hline \hline
	\textbf{100 modes}  & $2.97 \pm 1.17$ & $7.74 \times 10^{-6}$ & $2.03 \times 10^{-3}$\\
	\textbf{ROM2} & $8.02 \pm 2.19$ & $2.11 \times 10^{-6}$ & $1.49 \times 10^{-3}$\\
	\textbf{ROM1} & $12.98 \pm 3.46$ & $4.35 \times 10^{-6}$ & $1.89 \times 10^{-3}$\\
	\hline
	\end{tabular}
	\caption{ Reduced-order models comparison based on the LSTM-1-200-10-100 architecture. All the 100 modes are considered in the reconstruction. Data is averaged over three different trainings performed with the same parameters.}
	\label{LSTM_stats_RoMs}
\end{table}
\Cref{LSTM_stats_RoMs} shows how the error on the statistics increases as fewer modes are employed for training. The variance associated with the different trained models follows this same trend. Note that, the losses are smaller when reducing the number of modes which enter the training process, because the network has less data to validate.

\subsection{Reduced models based on an alternative field reconstruction}
\label{sec:ROMs_b}

\begin{figure}[h]
    \centering
    \includegraphics[width=4.6in]{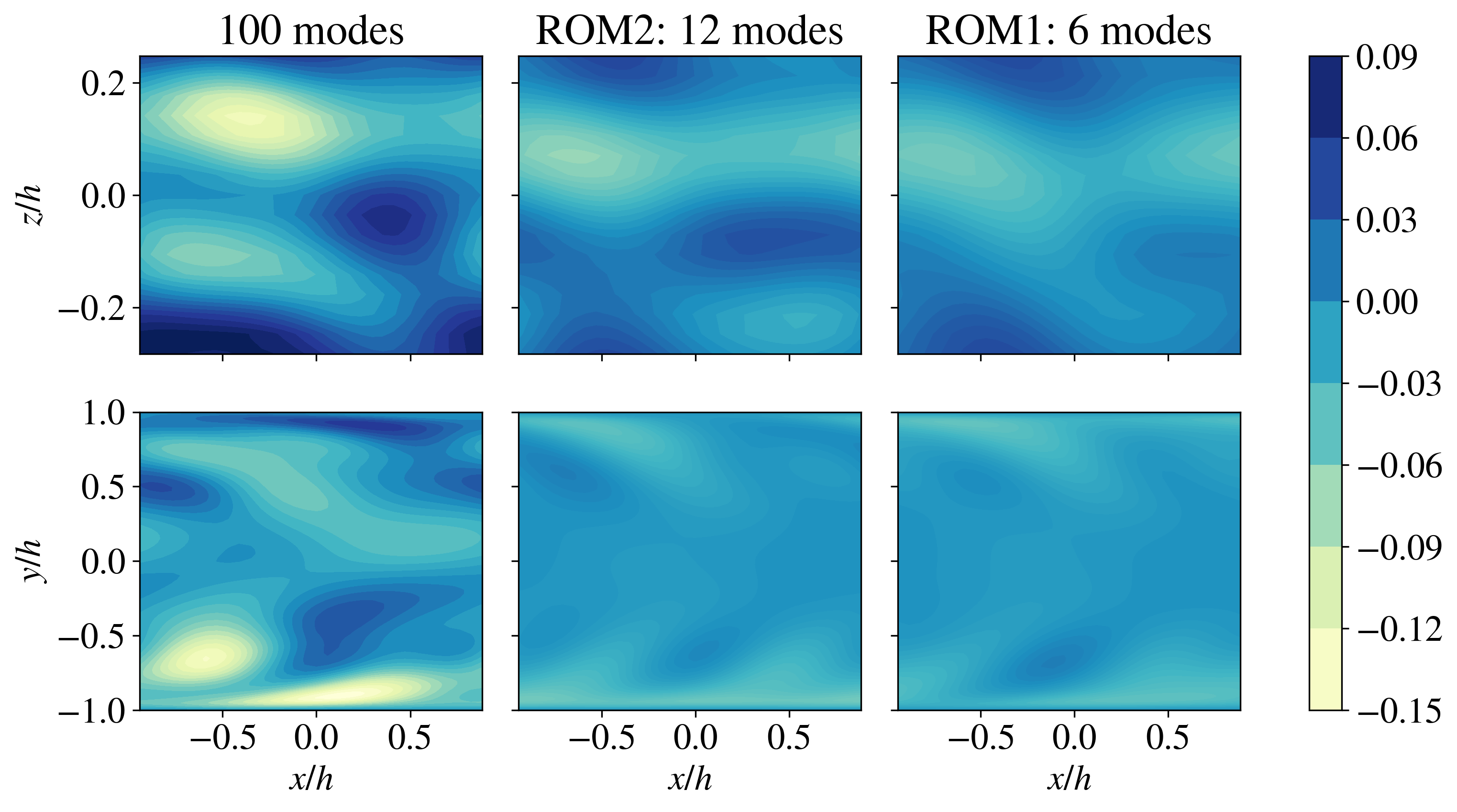}
    \caption{Instantaneous reconstruction of the streamwise velocity fluctuations with: the first 100 most energetic modes (\emph{left}), 12 most energetic modes of each sub-group (ROM2 - \emph{center}) and 6 most energetic modes of each sub-group (ROM1 - \emph{right}). Reconstruction at $t = 10,000$, at (\emph{top}) $z = 0$ and (\emph{bottom}) $y^+ \approx 24$ (distance from the wall in viscous units).}
    \label{ROMs_compare_visual}
\end{figure}

\begin{figure}[h]
    \centering
    \includegraphics[width=4.5in]{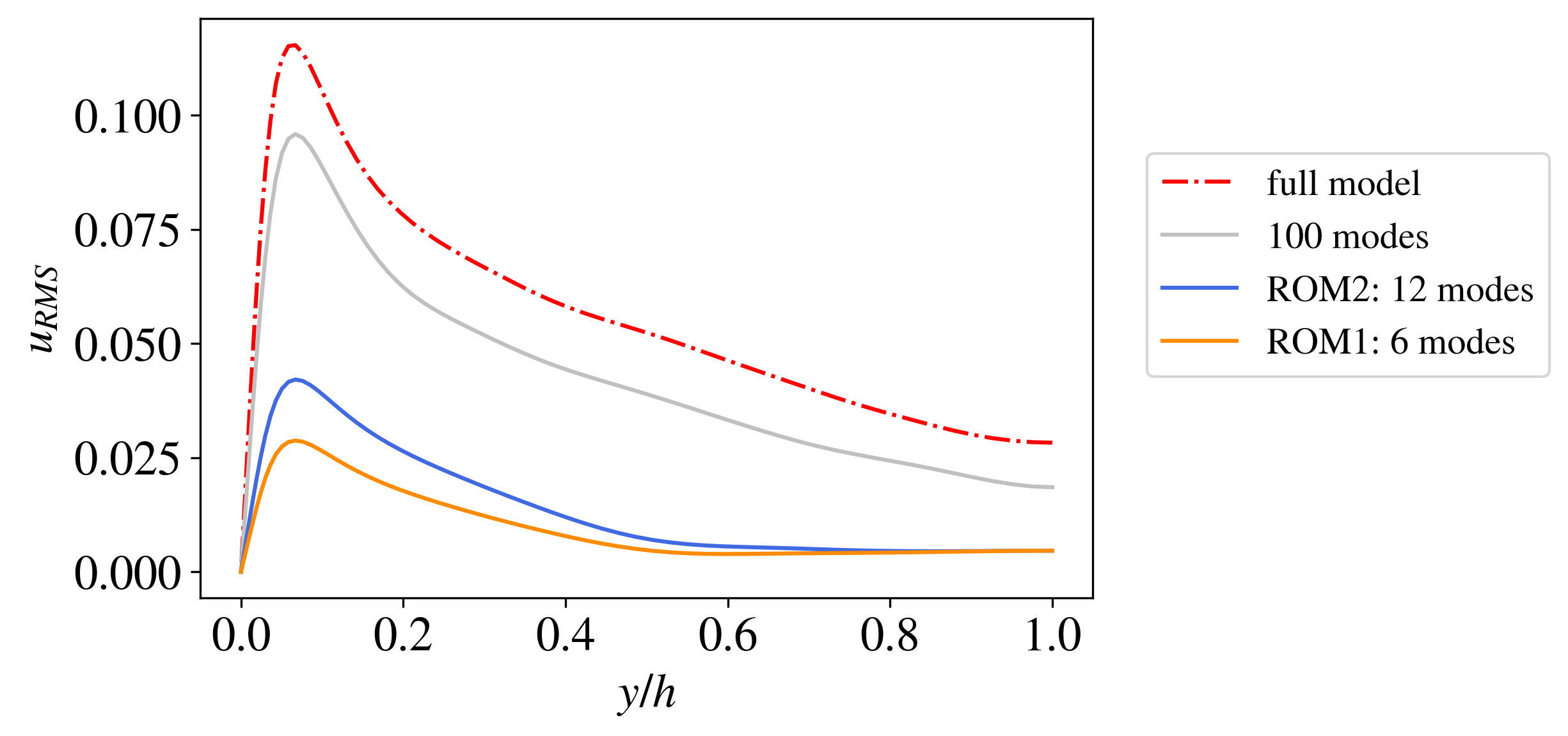}
    \caption{Comparison of the true $u_{RMS}$ profiles when reconstructing the field with all the modes (\emph{red}), with respect to the data-driven case (100 modes, \emph{grey}) and the case with 12 and 6 modes (\emph{blue} and \emph{orange}, respectively).}
    \label{ROMs_urms_compare}
\end{figure}

This approach employs the models developed in the previous section and due to this the validation losses are the same ones as in \cref{LSTM_stats_RoMs}. 
The difference lies in the reconstruction of the velocity fields, where here only one or two modes for each turbulent feature are employed. As illustrated in \cref{ROMs_compare_visual}, the intensity of the streamwise fluctuations is reduced, leading to a less detailed reconstruction, which is still able to incorporate the most relevant structures necessary to describe the minimal-channel turbulence.
This is also highlighted in \cref{ROMs_urms_compare}, where the profiles of the stream-wise fluctuations are reported for the full-model case, \emph{i.e.} the original velocity fields employing all the $n_x \times n_y \times n_z = 66048$ modes, where we consider the number of grid points having $e = n_y$ as the size of the economy-SVD which has to be performed for each wavenumber pair, thus $n_x \times n_z$ times; the data-driven model, where 100 modes are considered, the ROM2 (with 12 modes) and the ROM1 (with 6 modes).
The reduction in the reconstructed fluctuations is connected with a smaller fraction of energy captured by the alternative reconstructions, indeed we are able to describe approximately only the $10 \%$ or $20 \%$ of the total energy of the fluctuations $\mathbf{u}'(\mathbf{x},t)$ with respect to the $97 \%$ reconstructed with the 100 modes of the data-driven model, where also degeneracy is accounted for (\cref{LSTM_RoMs_alt_table}). 
\begin{table}[h]
\centering
	\begin{tabular}{| c | c c |}
	\hline \multirow{2}{*}{Model} & Reconstructed & \multirow{2}{*}{$E_{u_{\rm{RMS}}}[\%]$}  \\
	& energy $\%$ & \\
	\hline \hline
	\textbf{100 modes} & 97.05 & $2.97 \pm 1.17$ \\
	\textbf{ROM2: 12 modes} & 20.71 & $6.69 \pm 3.52$\\
    \textbf{ROM1: 6 modes} & 10.74 & $8.51 \pm 5.06$ \\
	\hline
	\end{tabular}
	\caption{Alternative ROMs: energy percentage (on fluctuations only) and errors with respect to the reference $u_{RMS}$ profile are reported. In the two ROMs the degeneracy is not considered.}
	\label{LSTM_RoMs_alt_table}
\end{table}

\begin{figure}[h]
    \centering
    \includegraphics[width=4.6in]{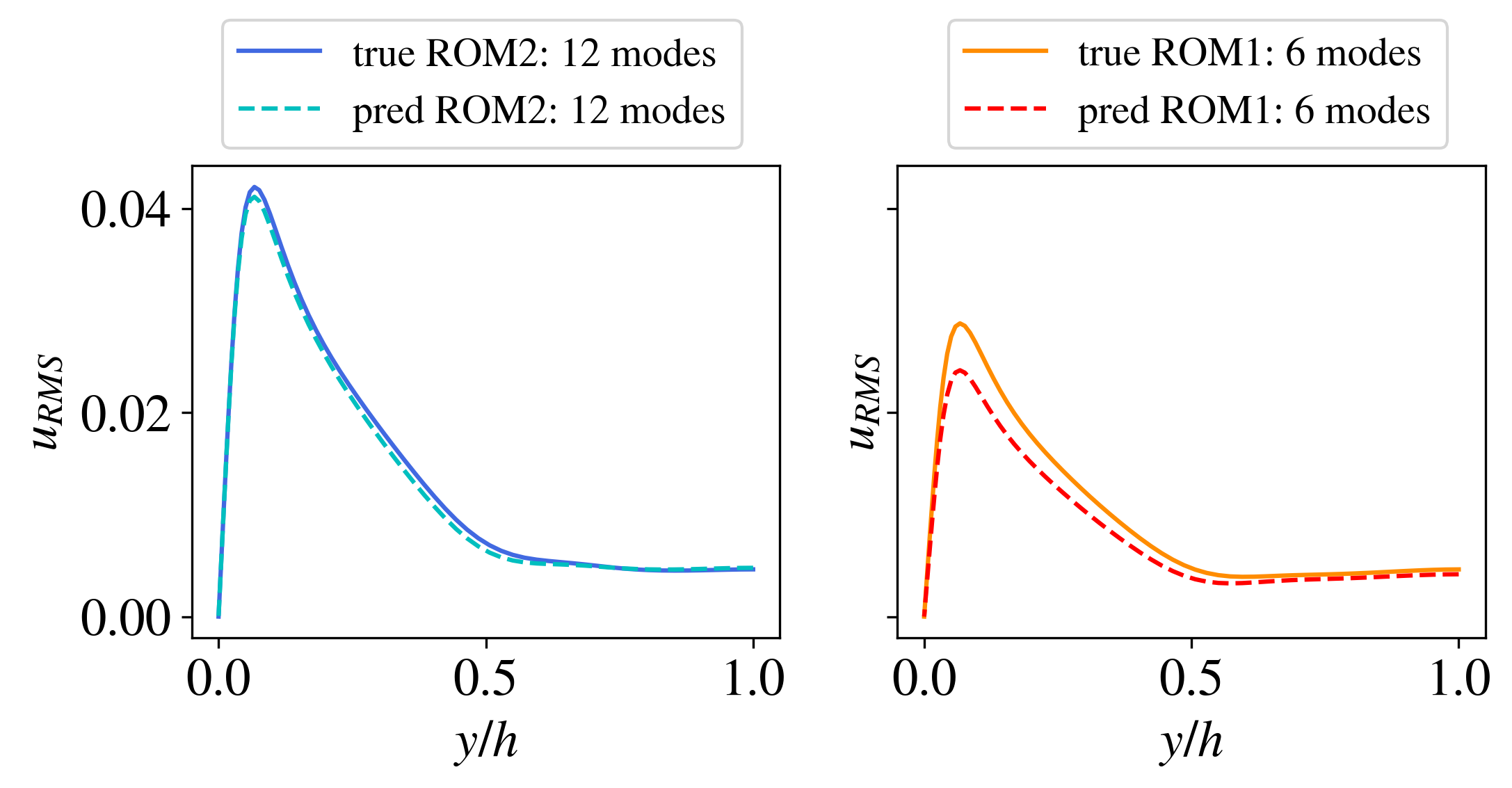}
    \caption{Streamwise velocity fluctuation profiles for the ROM2: 12 modes (\emph{left}) and the ROM1: 6 modes (\emph{right}) alternatives for the best models. The solid line represents the true profile and the dashed shapes the predicted ones.}
    \label{ROMsalt_urms_pred}
\end{figure}

Having the velocity reconstructions which capture a different energy percentage, statistics are compared separately for each one of the alternative models. Once again three different trainings are performed with the same setting to account for the stochastic nature of this process and the average of the errors over the fluctuations is shown in \cref{LSTM_RoMs_alt_table} together with their relative standard deviations. In this case we also observe that the errors and the variance increase as the number of modes which enter the computations is reduced.
\Cref{ROMsalt_urms_pred} shows the predictions of the streamwise velocity fluctuation profiles for the ROM2 and ROM1 cases. The true profiles are given by the solid line and the dots show the predicted ones. We have an error of $3.3 \%$ with 12 modes included in the reconstruction and $4.7 \%$ with 6 modes, whereas $1.3 \%$ was the error given by the best model based on a 100-mode energy truncation.

\section{Summary and conclusions}
\label{sec:discussion}

With this analysis we want to explore the capabilities of neural networks to reproduce the chaotic motion of a minimal channel flow. Our previous studies~\citep{srinivasan, Eivazi_koopman} have shown how excellent predictions of the temporal dynamics are achievable for a low-order model of near-wall turbulence~\citep{mohelis} by using either long-short-term-memory (LSTM) networks or a Koopman-based framework where non-linearities are modelled through a forcing term (KNF). Here, we consider a data-driven model extracted from the outputs of a modal decomposition over which an energy-based truncation is performed to define the number of modes that are needed to describe the minimal channel turbulence in an exhaustive way. The decomposition has been achieved through a proper-orthogonal decomposition in the Fourier domain (FFT-POD), already employed in studies such as the one from \cite{webber_sirovich}. The effectiveness and the benefits of this method have been demonstrated from the computational standpoint, with respect to the POD in the physical domain.
The simplified model which enters the computation accounts for the first 100 most energetic modes and captures nearly $80 \%$ of the total energy of the fluctuations in the $x$-direction. The spatial basis associated with the same model also include the most relevant turbulent features which describe the mean flow and its evolution, the streamwise counter-rotating vortices and the tilted vortices, with a dependence in the three directions, which are responsible for the transport of the vorticity.
The analysis of the temporal modes has spotlighted the necessity of a multi-step model for the training of the network, having different groups of signal which are characterized by a different frequency content.
A final simplification has been obtained by focusing only on the fluctuations not related to wavenumber pair (0,0).

The LSTM network outperforms the KNF, showing an excellent ability to predict the statistical behaviour of the turbulent flow. The best model employs a single layer of 200 neurons with an initial sequence of $p=10$ and is trained on a dataset comprising $100,000$ snapshots (LSTM-1-200-10-100). Predictions with this network result in an error on the streamwise fluctuations of $E_{u_{\rm{RMS}}} = 1.3\%$. Validation losses are of the order of $10^{-6}$ and $10^{-3}$ for the low and high frequencies. The instantaneous evolution of the predicted coefficients is in the correct range and it reproduces accurately the frequency content. Moreover, the quality of the predicted physics has been assessed through the Lyapunov exponent and the Poincaré maps, showing the capability of LSTM to reproduce the chaotic nature of the minimal channel turbulence and its dynamic behaviour. 
The limitations related to the KNF method have been instead highlighted given the data-driven nature of the problem, which makes the dynamics of the system significantly more complex when compared to the simplified case described by the model by \cite{mohelis}.
We also explored the possibility of implementing several reduced-order models (ROMs). The main idea is to employ a reduced amount of information based on the identification of turbulent structures in order to have a more efficient training. The LSTM network is able to reproduce the statistics associated with the reduced models, even if the accuracy is reduced as fewer modes are accounted during the training.

The present investigation can be extended in different directions: first, the POD analysis could be extended over the three velocity components, so that the complete description of the flow is available. This allows to assess whether the LSTM network is able to derive accurate statistics also in the spanwise and wall-normal directions and to study quantities as the Reynolds shear stress $\langle u'v' \rangle$, for instance. The net-flux modes, \emph{i.e.} fluctuations associated with wave-number pair (0,0), can also be included in the training and predictions.
There is still a margin of improvement for the step-by-step predictions by either employing new combinations of the hyper-parameters for the LSTM network or by elaborating a more complex KNF procedure to handle this case. 
The prediction can also be improved, introducing a more sophisticated network which includes a loss function based on both the instantaneous and statistical behaviour, for instance.
Future work will aim at other canonical wall-bounded flows, as pipe and duct flows, open channels~\citep{guastoni_CNN} or boundary layers.

\section*{Acknowledgments}

RV acknowledges the financial support from the G\"oran Gustafsson foundation and the Swedish e-Science Research Centre (SeRC). Part of the analysis was performed on resources provided by the Swedish National Infrastructure for Computing (SNIC) at PDC and HPC2N.

\bibliographystyle{elsarticle-harv}
\biboptions{authoryear}
\bibliography{References}

\end{document}